\documentclass[12pt]{iopart}

\usepackage{iopams}  
\usepackage{setstack, amsMatrix}

\usepackage{bm}
\usepackage{amssymb}
\usepackage{amsthm}
\usepackage{graphicx}
\usepackage{subfigure}
\usepackage{rotating}
\usepackage{mathrsfs}
\usepackage{wasysym}
\usepackage{stmaryrd}

\newcommand{ \ket }[1]{ | #1 \rangle }

\begin{document}

\title[Why momentum width matters for atom interferometry]{Why momentum width matters for atom interferometry with Bragg pulses}

\author{S. S. Szigeti, J. E. Debs, J. J. Hope, N. P. Robins, and J. D. Close}

\address{Department of Quantum Science, Research School of Physics and Engineering, The Australian National University, Canberra, 0200, Australia}

\ead{stuart.szigeti@anu.edu.au}

\begin{abstract}
We theoretically consider the effect of the atomic source's momentum width on the efficiency of Bragg mirrors and beamsplitters and, more generally, on the phase sensitivity of Bragg pulse atom interferometers. By numerical optimization, we show that an atomic cloud's momentum width places a \emph{fundamental} upper bound on the maximum transfer efficiency of a Bragg mirror pulse, and furthermore limits the phase sensitivity of a Bragg pulse atom interferometer. We quantify these momentum width effects, and precisely compute how mirror efficiencies and interferometer phase sensitivities vary as functions of Bragg order and source type. Our results and methodology allow for an efficient optimization of Bragg pulses and the comparison of different atomic sources, and will help in the design of large momentum transfer Bragg mirrors and beamsplitters for use in atom-based inertial sensors.
\end{abstract}

\pacs{03.75.Be, 03.75.Dg, 37.25.+k, 67.85.Hj}
\vspace{2pc}
\noindent{\it Keywords}: Atom optics, Atom interferometry, Bragg diffraction, Bose-Einstein condensation

\submitto{\NJP}
\maketitle

%%%%%%%%%%%%%%%%%%%%%%%%%%  body  %%%%%%%%%%%%%%%%%%%%%%%%%%
\section{Introduction}
The development of large momentum transfer (LMT) mirrors and beamsplitters for atoms is a promising path for increasing precision in future sensors based on atom interferometry \cite{McGuirk:2000, Muller:2009,Clade:2009}. In particular, Bragg diffraction of atoms from a light pulse seems to be a strong candidate for LMT atom-optical elements, and has received much theoretical \cite{Giltner:1995b, Blakie:2000, Muller:2008, Benton:2011} and experimental attention \cite{Giltner:1995, Kozuma:1999, Hughes:2007, Muller:2008b,Debs:2011}. However, much theoretical work on Bragg diffraction assumes a plane-wave atomic source. In this paper, we relax this assumption, and theoretically investigate how the momentum width of an atomic source affects the efficiency of Bragg beamsplitters and mirrors and, more generally, the signal-to-noise ratio (SNR) of a Mach-Zehnder (MZ) interferometer. In particular, we demonstrate that the momentum width places fundamental restrictions on the Bragg mirror efficiency, and consequently limitations on the interferometer sensitivity. In addition, limitations on laser power and spontaneous loss place further restrictions on efficiency and sensitivity that can only be overcome by severely narrowing the momentum width. The analysis presented in this paper gives one the means of optimizing Bragg mirror and beamsplitter pulses, and furthermore allows one to determine the source type, momentum width and momentum transfer that maximize the signal-to-noise ratio of a Bragg pulse atom interferometer.

To date, cold atom inertial sensing has been dominated by thermal sources typically at a temperature on the order of $1\,\mu$K  or less. Whilst such sources can be created with a high atom flux, they typically have a large momentum width on the order of $1 \,\hbar k$, where $k = |\mathbf{k}|$ is the wavenumber of the laser light. In experimental work on LMT mirrors and beamsplitters carried out by a number of groups, it has been necessary to velocity select the source cloud (reducing the atom flux) prior to beamsplitting in order to realize high fidelity beamsplitting and reflection \cite{McGuirk:2000, Muller:2008b, Peters:2001}. This demonstrates that, practically, a narrower momentum width is preferable even though it leads to a reduction in total atom flux. 

There has been some interest in understanding whether it is advantageous to use a narrow momentum source, such as a Bose-Einstein condensate (BEC), for atom interferometry. For example, Kozuma \emph{et al.} \cite{Kozuma:1999} and Torii \emph{et al.} \cite{Torii:2000} have used low order Bragg diffraction to transfer a condensate to a target momentum state with an efficiency close to $100 \%$. Hughes \emph{et al.} \cite{Hughes:2007} considered how the fidelity of Bragg beamsplitters and mirrors depends upon the atomic velocity of a BEC, and hence showed that the narrow momentum distribution of a BEC allowed for very efficient 1st, 2nd and 3rd order Bragg beamsplitters and mirrors. More recently, the phase sensitivity of the atom interferometer developed by Impens \emph{et al.} \cite{Impens:2009, Impens:2011}, which involves the levitation of a condensate, has been shown to improve when the atomic source is progressively narrowed through filtering with multiple wave atomic interferences. Debs \emph{et al.} \cite{Debs:2011} constructed a BEC gravimeter, utilizing Bragg LMT mirrors and beamsplitters, that had superior fringe contrast to a similar thermal source gravimeter. 

Whilst such results illustrate the importance of a narrow momentum width source, one must recall that current Bose-condensed sources have a lower atom flux than state-of-the-art thermal sources. Indeed, a state-of-the-art BEC machine that produces $2.4 \times 10^6$ atoms/s \cite{van_der_Stam:2007} still has a flux $25$ times lower than the best published state- and velocity-selected thermal cloud used in a gravimeter \cite{Muller:2008c}. It has thus been unclear whether the advantages of a narrower momentum width compensate for the reduced atom flux. We quantitatively address this question, and accommodate both flux and momentum width considerations by optimizing the only parameter that matters: the signal-to-noise ratio.

Previous experimental and theoretical work has also suggested that as the order of the Bragg process is increased (which increases the momentum imparted during the process), a narrower momentum width is required in order to realize high efficiency beamsplitting and mirror processes. For example Catallioti \emph{et al.} \cite{Cataliotti:2001} reported experimental and theoretical results indicating that the linewidth of multiphoton transitions decreases as the reciprocal of the order of the process, at least in the perturbative regime where the excited state population of the two-level system is kept small. For a condensate, Wu \emph{et al.} \cite{Wu:2005} showed that the transfer efficiency of a particular two-pulse Bragg beamsplitter decreased as the Bragg order increased. Interestingly, our analysis shows that the maximum possible transfer efficiency of Bragg mirrors and beamsplitters does \emph{not} degrade at higher Bragg orders for clouds of momentum width less than approximately 0.25 $\hbar k$, on the proviso that one is not limited by laser power. Whilst the maximum transfer efficiences of larger momentum width clouds do vary with Bragg order, this dependence is small and is almost negligible for $n \geq 5$. However, when the two-photon Rabi frequency is restricted our analysis reveals that the efficiency of higher order Bragg processes is reduced by a factor that strongly depends upon the momentum width of the cloud. As the Bragg order increases, this degradation in efficiency becomes more severe and can only be overcome by increasingly narrow momentum widths and longer pulse durations. This demonstrates that the practical implementation of high fidelity LMT Bragg mirrors and beamsplitters necessarily requires a narrow momentum width cloud and a large laser intensity.

Our paper is structured as follows. \textbf{Section~\ref{Background}} briefly introduces a MZ atom interferometer and motivates the interest in LMT for atom interferometry.  \textbf{Section~\ref{Theoretical_model}} introduces a model that describes Bragg diffraction, and develops a criterion that can be used to optimize Bragg mirror pulses. In \textbf{section~\ref{Mirror_numerics}}, numerical simulations using this model and criterion demonstrate that: (a) the momentum width places a \emph{fundamental} limit on the reflection efficiency of a Bragg mirror; (b) for unconstrained two-photon Rabi frequency, the maximum efficiency of a mirror pulse does \emph{not} markedly change with increasing Bragg order; and (c) the optimal two-photon Rabi frequency scales roughly quadratically with the order of the Bragg process. Qualitative features of this numerical analysis are explained using a simple two-level model in \textbf{section~\ref{Sec_2_level}}. In \textbf{section~\ref{mirror_laser_limit}}, it is shown that when the two-photon Rabi frequency is limited (due to limitations on laser power and spontaneous loss), (a) the efficiency of a Bragg mirror drastically decreases above some Bragg order; and (b) achieving the same mirror efficiency at a higher Bragg order requires a reduction in the momentum width. The latter half of the paper is concerned with performing a similar analysis for a MZ interferometer constructed with Bragg mirrors and beamsplitters. Specifically, \textbf{section~\ref{Efficiency_MZ}} constructs an efficiency criterion for a MZ interferometer, thus allowing one to optimize both Bragg beamsplitters and mirrors, \textbf{section~\ref{2_level_MZ}} contains a qualitative analysis of a MZ interferometer with the two-level model, and \textbf{section~\ref{Results_MZ}} presents the quantitative results of the numerical optimization, with and without constraints on the two-photon Rabi frequency. The results mimic that of the Bragg mirror, showing that the momentum width fundamentally limits the maximum SNR for a phase measurement with a MZ interferometer, and particularly so when the two-photon Rabi frequency is constrained. In addition, a constraint on the two-photon Rabi frequency implies that, above some Bragg order, the SNR actually \emph{worsens}. Finally, in \textbf{section~\ref{Sec_BEC_vs_thermal}} the optimal Bragg order and (for a thermal source) velocity selection that maximizes the SNR of a MZ interferometer are calculated. A comparison of current state-of-the-art thermal and Bose-condensed sources demonstrates that, in principle, current Bose-condensed sources are competitive with state-of-the-art thermal sources. It is argued that future increases in atomic flux and the practical advantages of a narrow momentum width source imply that Bose-condensed sources have a significant role to play in the next generation of precision sensors based on Bragg pulse atom interferometry.   

\section{Background} \label{Background}
A comprehensive review of atom interferometry is given in \cite{Cronin:2009}. We consider an atom interferometer in the Mach-Zehnder ($\pi/2-\pi-\pi/2$) configuration, where $n$th order Bragg transitions are used as beamsplitters ($\pi/2$ pulse) and mirrors ($\pi$ pulse). Such an interferometer operates by coupling two momentum states $\ket{p_0}$ and $\ket{p_0+2n \hbar k}$, where $k = |\mathbf{k}|$ is the wavenumber of the laser light and $n$ is the (integer) order of the Bragg process. If the acceleration of the atoms relative to the laser is uniform, then the free evolution phase in each arm is identical, and the only interferometric phase contribution is from the atom-light interaction. In particular, it is straightforward to show that the interferometric phase shift is proportional to the acceleration \cite{Kasevich:1992}:
\begin{equation}
	\Phi = n(\phi_1 - 2\phi_2 + \phi_3) = 2n\mathbf{k} \cdot \mathbf{a} T^2,
\end{equation}	
where $\phi_i$ is the optical phase of the $i$th Bragg pulse at the point where the atom interacts with the pulse, relative to some reference position, $\mathbf{a}$ is the uniform acceleration (due to gravity, say) relative to that reference position, and $T$ is the time between pulses. If we scan the phase $\Phi$ and measure the relative population in the momentum state $\ket{p_0 + 2n \hbar k}$, then we obtain a signal proportional to $N\left(1+ \mathcal{C} \cos\Phi\right)/2$, where $N$ is the atomic population in the two momentum states and $\mathcal{C}$ is the contrast of the fringes. Therefore, for operation at mid-fringe and small phase shift $\Phi$, the SNR at the shot noise limit is
\begin{equation}
	SNR \propto \mathcal{C}\sqrt{N} \left( 2n \hbar \mathbf{k} T^2 \right) \cdot \mathbf{a}. \label{SNR}
\end{equation}
The motivation for LMT with Bragg transitions is now clear: the signal can be increased in direct proportion to increasing the momentum transferred between the two states (i.e. increasing the Bragg order $n$). 

\begin{figure}[t]
\centering
\subfigure[Ideal Bragg mirror process (plane-wave input, effective two-level system)]{
\includegraphics[width=\textwidth]{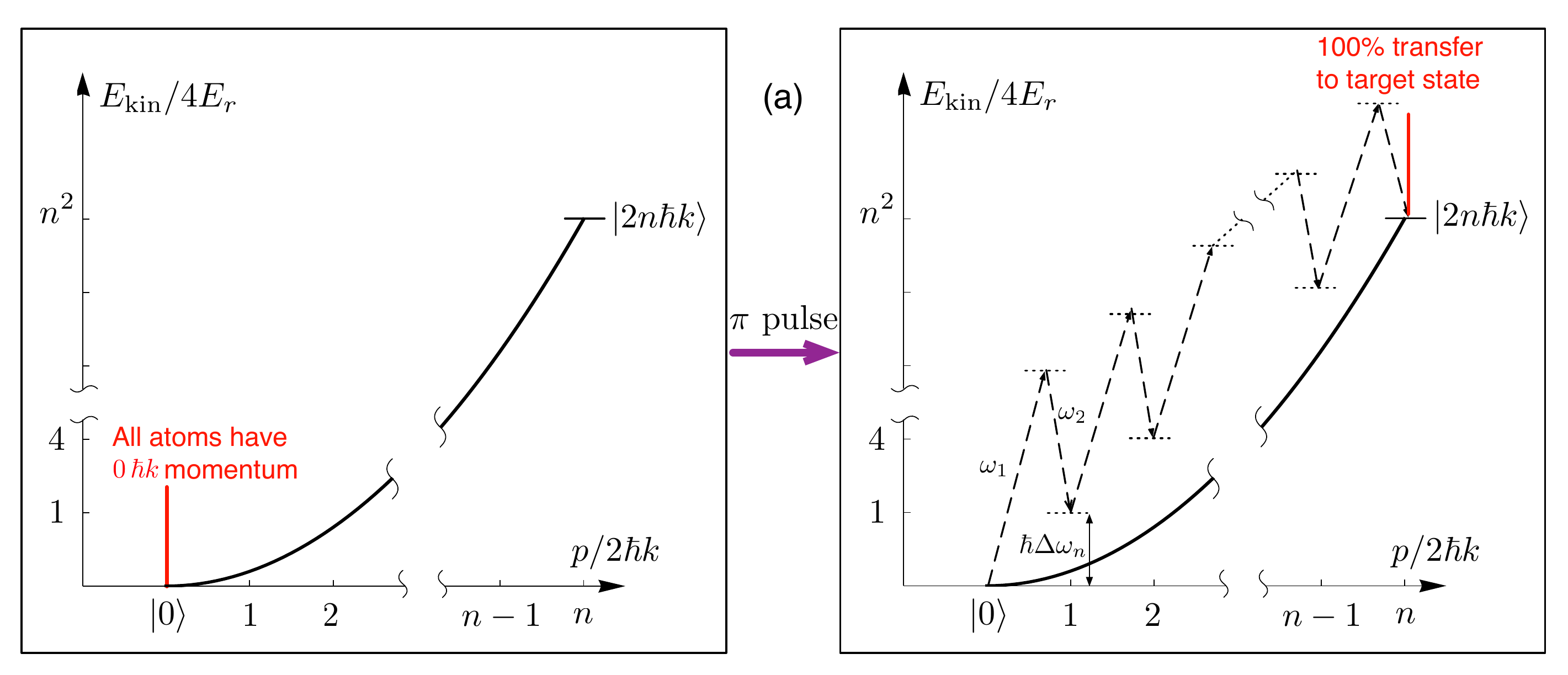}
\label{Ideal_Bragg}
}
\subfigure[Realistic mirror process (general initial momentum distribution, multi-level system)]{
\includegraphics[width=\textwidth]{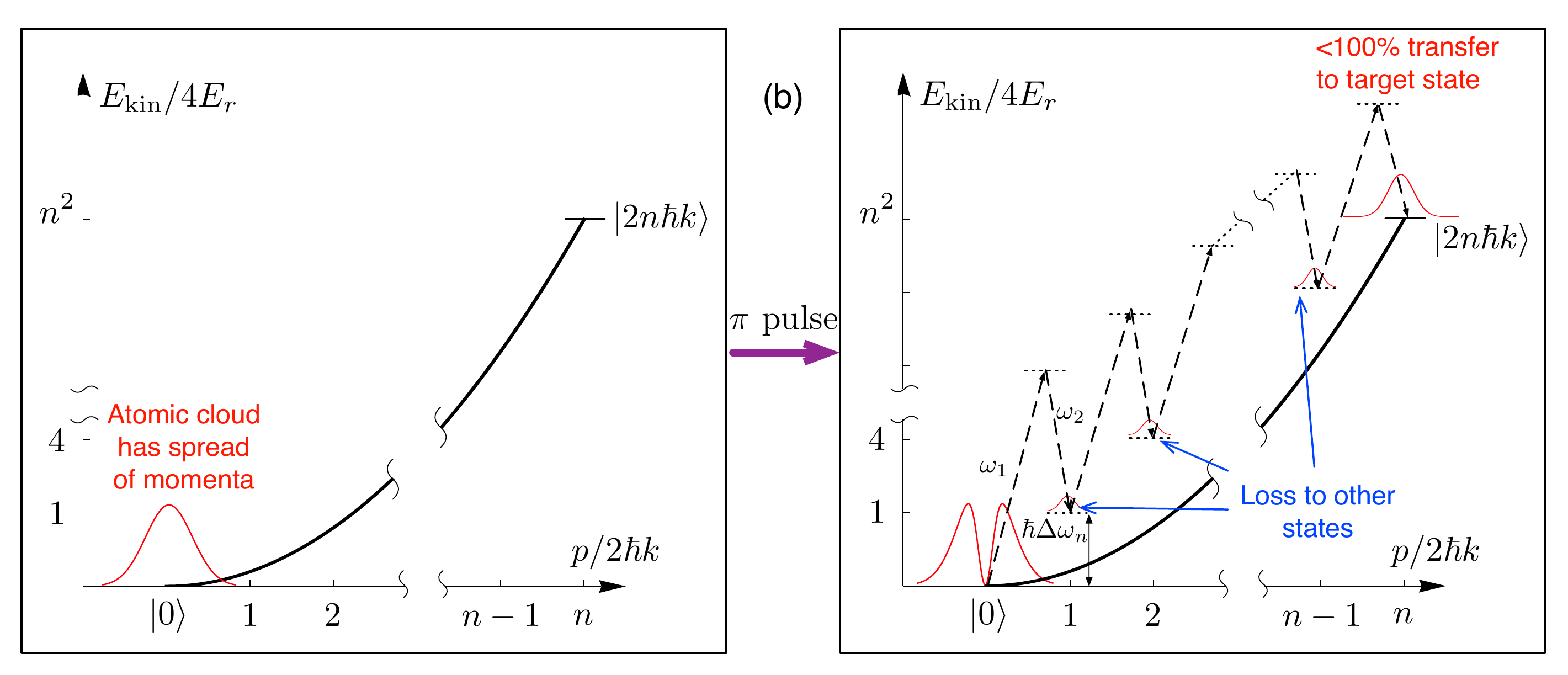}
\label{Real_Bragg}
}
\label{Bragg_diagram}
\caption{Energy $E_ \mathrm{kin}$ vs momentum $p$ diagrams illustrating the difference between an ideal Bragg mirror process and a realistic diffraction process. Here $E_r = \hbar \omega_r$ is the recoil energy. Without loss of generality, these diagrams show the case where $p_0 = 0$. An ideal Bragg process is shown in (a). In the realistic process (b), the initial atomic cloud has a finite momentum width. In this case it cannot be transferred to the target state with $100\%$ efficiency, and furthermore there is loss to diffraction orders other than the two interferometer modes.}
\end{figure}
In an ideal Bragg process, an atom in a single momentum state $\ket{p_0}$ is coupled to a final momentum state $\ket{p_0+2n\hbar k}$. Specifically, for $n$th order diffraction, $n$ photons at frequency $\omega_1$ and wave vector $\textbf{k}$ are absorbed and stimulated to emit into frequency $\omega_2$ with wavevector $-\textbf{k}$, leading to a transfer of momentum $2n \hbar k$ and an energy of $n \hbar \Delta \omega_n$ [cf. figure~\ref{Ideal_Bragg}]. However, as shown in figure~\ref{Real_Bragg}, in a realistic system the atomic cloud will initially be in some momentum distribution of width $\sigma$. Hence we cannot expect that, in general, perfect $\pi$ and $\pi/2$ pulses exist. Furthermore, realistic diffraction processes couple the initial momentum state $\ket{p_0}$ to a range of momentum states. This results in a loss of flux from the interferometer, which we would like to minimize. Indeed, as shown by M\"uller \emph{et al.}  \cite{Muller:2008}, practical limitations on pulse duration \emph{require} operation in the multi-level quasi-Bragg regime, which is the intermediate regime between the long interaction Bragg regime and the short interaction Raman-Nath regime. The theoretical model introduced in the next section places no restriction on the momentum width of the atomic cloud, nor the number of momentum states that take part in the diffraction process.

\section{Theoretical model} \label{Theoretical_model}
We consider a semiclassical model of Bragg scattering, where a cloud of two-level atoms interacts with a classical light field. This model has been successfully applied to similar Bragg scattering problems \cite{Meystre:2001}, and is therefore suitable for our purposes. Whilst figure~\ref{Bragg_diagram} shows Bragg diffraction in the frame where a stationary atom interacts with a moving optical lattice, it is more convenient to work in the frame where the lattice is stationary. In this case, an $n$th order Bragg resonance corresponds to a resonant atomic momentum of $- n \hbar k$, for integer $n$. At this point, we assume an initial momentum state of $(-n+\delta)\hbar k$.  Here $\delta \hbar k$ can be thought of as the momentum detuning from resonant momentum. Provided the laser is far detuned from the excited state, the atomic excited state can be adiabatically eliminated as shown in \cite{Muller:2008}. In this limit, the dynamics are completely described by the Schr\"odinger equation for the ground state wavefunction $g(z,\delta, t)$:
\begin{equation}
	i \hbar \dot{g}(z,\delta, t) = - \frac{\hbar^2}{2 M} \frac{\partial^2 g(z,\delta, t)}{\partial z^2} + 2\hbar \Omega(t) \cos^2(kz)g(z,\delta, t), \label{g_eq}
\end{equation}
where $M$ is the mass of the atom, $z$ is the position co-ordinate along the direction of the standing wave and $\Omega(t)$ is the two-photon Rabi frequency. We have assumed that the atom-atom interaction energy is negligible and can be ignored. This is true for most atom interferometer experiments, including those using expanded BEC sources \cite{Debs:2011, Jamison:2011}, where these interactions would give an undesired systematic effect. 

Since the light potential is periodic, we apply Bloch's theorem and express $g(z,\delta, t)$ in terms of eigenfunctions of constant quasimomentum:
\begin{equation} 
	g(z,\delta,t) = \sum_{m = -\infty}^{\infty} c_{2m}(\delta, t)e^{i (2m+\delta) k z}. \label{Bloch_expansion}
\end{equation}
Substituting (\ref{Bloch_expansion}) into (\ref{g_eq}) gives the following set of ordinary differential equations (indexed by integer $m$):
\begin{equation}
	i \dot{c}_m(\delta,t) = \omega_r (m + \delta)^2 c_m(\delta,t) + \frac{1}{2}\left[\Omega(t)c_{m+2}(\delta,t) + \Omega^*(t)c_{m-2}(\delta,t)\right], \label{c_eq}
\end{equation}
where $\omega_r = \hbar k^2/2M$ is the atomic recoil frequency. For the 780 nm D2 transition of $^{87}$Rb, $\omega_r \sim 2\pi \times 3.8$ kHz. 

In what follows, we will consider Gaussian pulses, as they give lower losses to other diffraction orders than square pulses \cite{Muller:2008}. For a single pulse,
\begin{equation}
	\Omega(t) = \Omega \exp\left( -t^2/2\tau^2\right),
\end{equation}
where $\Omega$ is the amplitude of the pulse and $\tau$ is the pulse duration. For a MZ interferometer, there are three Gaussian pulses, separated by time $T$, and the phase difference between the two arms is included by introducing a phase shift $\phi$ on the third pulse:
\begin{eqnarray}
	\Omega(t) 	&= \Omega_\mathrm{bs}\exp\left( -t^2/2\tau_\mathrm{bs}^2\right)+\Omega_\mathrm{m}\exp\left( -(t+T)^2/2\tau_\mathrm{m}^2\right) \nonumber \\
				&\qquad +\Omega_\mathrm{bs}\exp\left( -(t+2T)^2/2\tau_\mathrm{bs}^2+i\phi\right). \label{3_pulse}
\end{eqnarray}
Here $\Omega_\mathrm{bs}$ $(\Omega_\mathrm{m})$ and $\tau_\mathrm{bs}$ $(\tau_\mathrm{m})$ are the pulse amplitude and duration for the beamsplitter (mirror), respectively. 

\subsection{Efficiency criterion for a mirror pulse}
Initially, we assume the atomic cloud to be in the state
\begin{equation}
	\ket{\psi_i} = \int  \rho(\delta) \ket{(-n+\delta)\hbar k}\, d\delta,
\end{equation}
where $\rho(\delta)$ is the cloud's momentum distribution initially centred at momentum $-n \hbar k$. We assume $\rho$ to be real, and note that it must satisfy $\int \rho(\delta)^2 \, d\delta= 1$ in order for $\ket{\psi_i}$ to be correctly normalised.

For an $n$th order Bragg transition a mirror pulse ideally maps $\ket{\psi_i}$ to the final state
\begin{equation}
\ket{\psi_\mathrm{Ideal}} = \int \rho(\delta) \ket{(n +\delta)\hbar k}\,d\delta . \label{psi_ideal}
\end{equation}
That is, a distribution $\rho$ of momentum states centred at momentum $-n \hbar k$ has been mapped to the \emph{same} distribution $\rho$ of momentum states centred at momentum $n \hbar k$. However, in general we will get a final state of the form
\begin{equation}
	\ket{\psi} = \sum_m \int  c_m (\delta,t_f) \ket{(m + \delta)\hbar k} \, d\delta, \label{psi_actual}
\end{equation}
where the coefficients $c_m(\delta,t)$ are determined by solving the set of equations (\ref{c_eq}), and $t_f$ is some time well after the pulse has been turned on and off (i.e. $\Omega(t_f) \approx 0$). We characterize the difference between the state we \emph{get} and the state we \emph{want} through the fidelity:
\begin{equation}
	F_{\pi} = |\left< \psi_\mathrm{Ideal} | \psi \right>|^2. \label{Fid_defn}
\end{equation}
By direct substitution of (\ref{psi_ideal}) and (\ref{psi_actual}) into (\ref{Fid_defn}), and application of the Cauchy-Schwartz inequality, we can prove that
\begin{equation}
	F_\pi \leq \int \rho(\delta)^2 \left|c_{n}(\delta,\tau_f)\right|^2 \, d\delta.
\end{equation}
If we assume that the relative phase between different momentum components of the cloud is approximately zero, then this upper bound will approximately be the fidelity. Thus, if we take $\rho(\delta)$ to be a normalised Gaussian of width $\sigma$, the efficiency criterion for a Bragg mirror is
\begin{equation}
	F_\pi \approx \int \frac{1}{\sqrt{2\pi \sigma^2}} e^{-\delta^2/ 2 \sigma^2} \left|c_{n}(\delta,\tau_f)\right|^2 \, d\delta. \label{F_pi_criterion}
\end{equation}
Given that $e^{-\delta^2/ 2 \sigma^2} \left|c_{n}(\delta,\tau_f)\right|^2/\sqrt{2\pi\sigma^2}$ is the population transferred to the state $\ket{(n+\delta)\hbar k}$, we can see that $F_\pi$ is just the total population that is transferred to a distribution $\rho(\delta)^2$ centred at momentum $n \hbar k$. Clearly, the best Bragg mirror will maximize transfer to this distribution, hence maximizing the fidelity. 

The efficiency criterion defines our optimization problem. Specifically, our goal is, for different Bragg orders $n$ and momentum widths $\sigma$, to find the optimal two-photon Rabi frequency $\Omega_\mathrm{opt}$ and pulse duration $\tau_\mathrm{opt}$ that maximize criterion (\ref{F_pi_criterion}).

\section{Analysis for a Bragg mirror} \label{Bragg_results}
\subsection{Numerical simulation with unconstrained power} \label{Mirror_numerics}
For different Bragg orders $n$ and momentum widths $\sigma$, the optimal pulse amplitude $\Omega_\mathrm{opt}$ and duration $\tau_\mathrm{opt}$ were found by maximizing the fidelity (\ref{F_pi_criterion}). $F_\pi$ was computed by numerically solving a truncated set of equations (\ref{c_eq}) for various $\delta, \Omega$ and $\tau$. Momentum states larger (smaller) than $m_\mathrm{max}$ ($-m_\mathrm{max}$) were truncated, as they were not populated during the diffraction process. Given that the initial and target states have kinetic energy $n^2 \hbar \omega_r $ and the potential energy due to the optical lattice is $\hbar \Omega$, the highest energy state $m$ to be populated is given by $n^2 \omega_r  + \Omega  \sim m^2 \omega_r $. Including the nearest two or three higher energy states to $m$, we thus set $m_\mathrm{max}$ to the integer part of $(\sqrt{\Omega/\omega_r+ n^2}+6)$ if $n$ was even, or to the integer part of $(\sqrt{\Omega/\omega_r+n^2}+7)$ if $n$ was odd.
 
For simplicity, we restricted the optimization for $\tau$ to be within the first Rabi cycle. This is not an unreasonable restriction, given that an interferometer requires pulse durations much shorter than the interrogation time (i.e. the time between successive pulses). 

\begin{figure}[ht!]
\centering
\subfigure{
\includegraphics[scale=0.6]{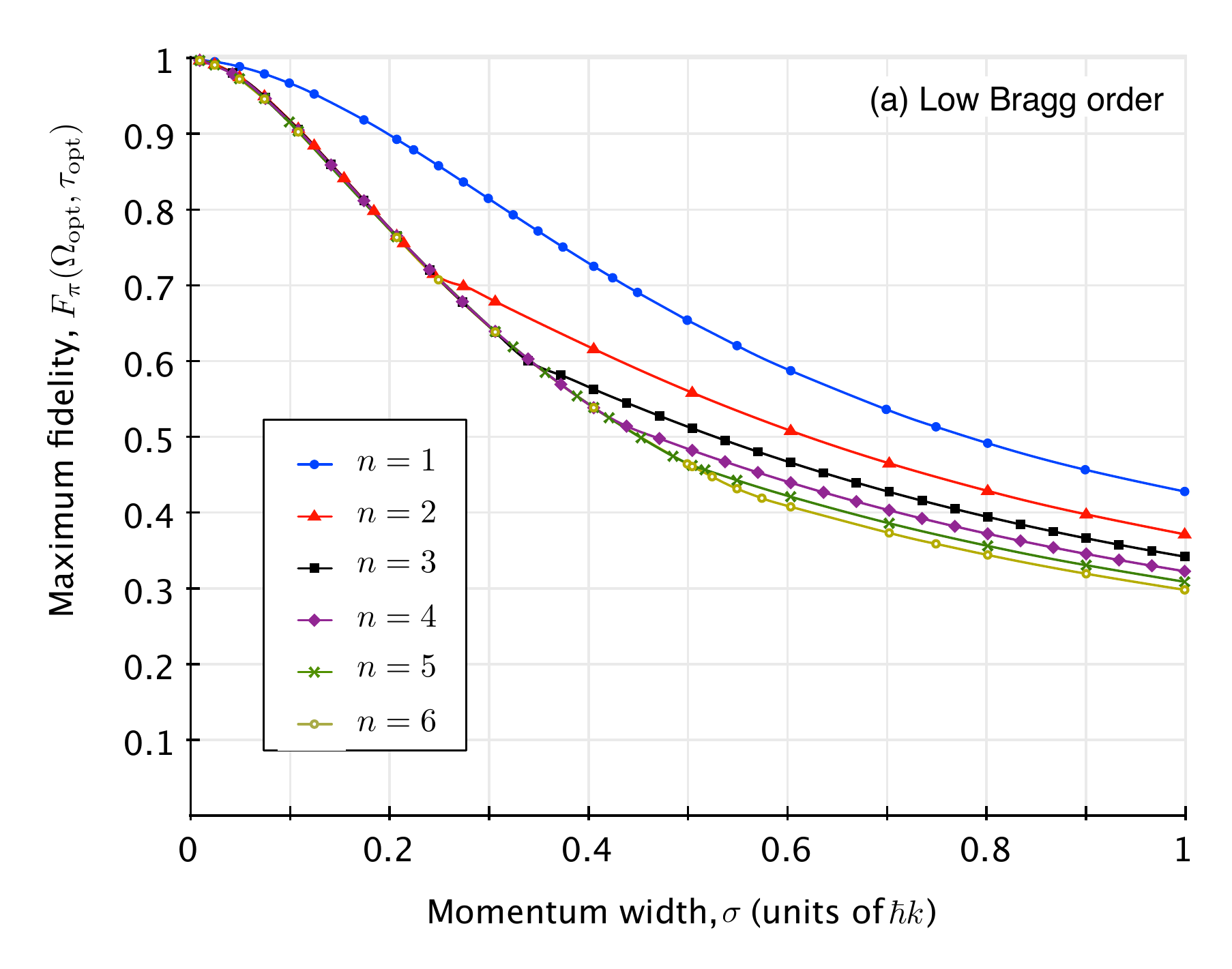}
}
\subfigure{
\includegraphics[scale=0.6]{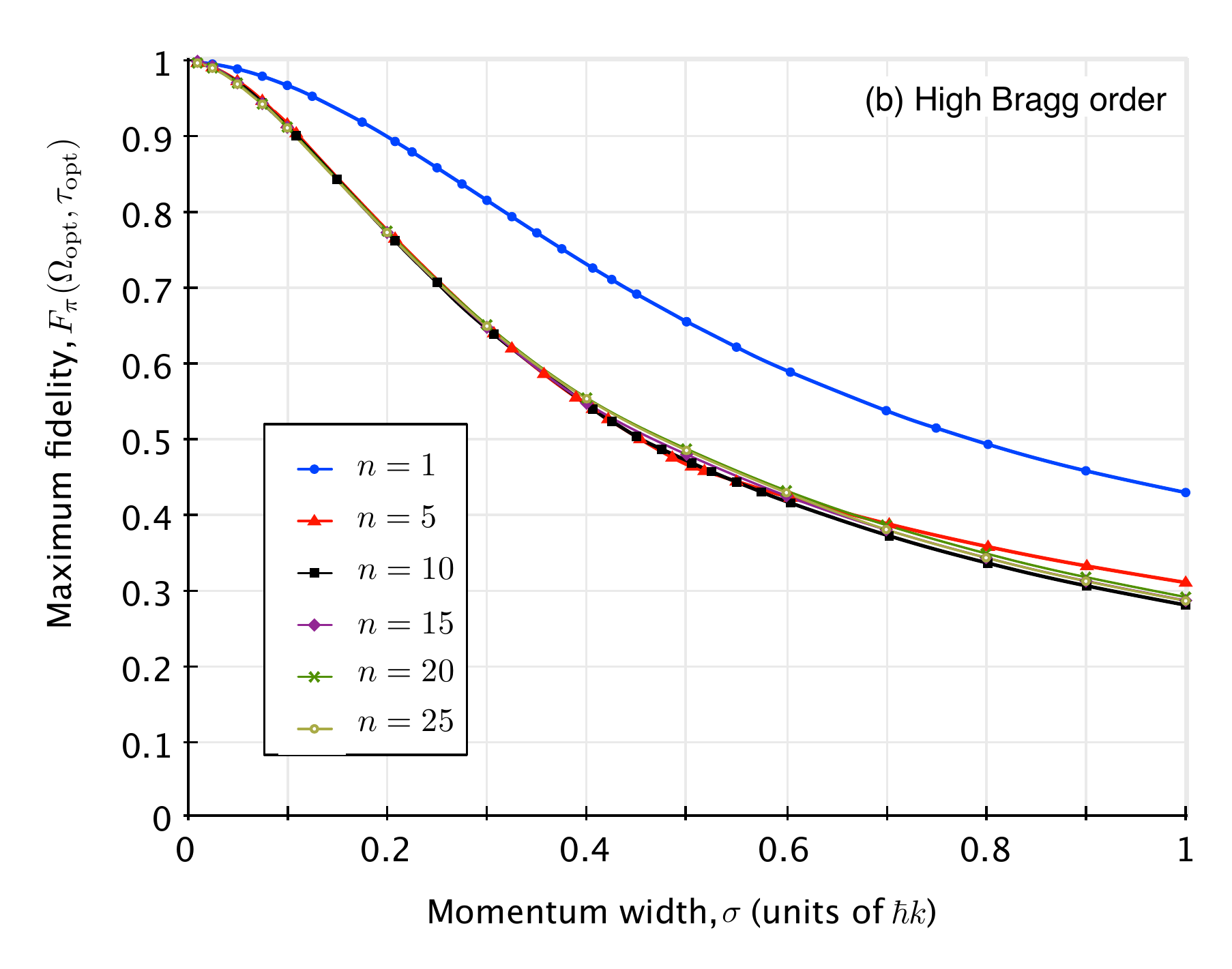}
}
\label{Fid_plots}
\caption{\label{Fid_plots} Plots of the maximum fidelity $F_\pi(\Omega_\mathrm{opt},\tau_\mathrm{opt})$ as a function of momentum width $\sigma$. Different curves correspond to different order Bragg processes. The curves have been separated into two plots to better allow for discrimination between them. Note that the actual data is given by the points; the lines are simply to guide the eye.}
\end{figure}
\begin{figure}[t!]
\centering
\includegraphics[scale=0.37]{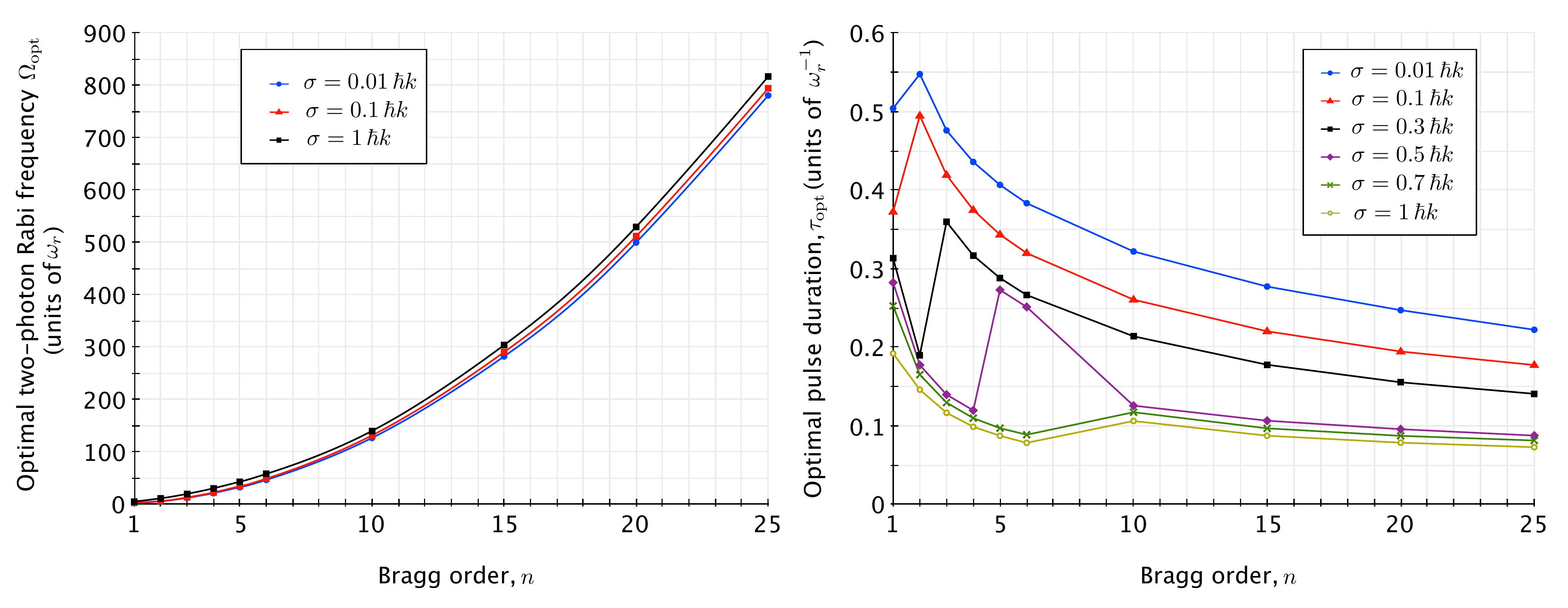}
\caption{\label{Mirror_opt_plots} Plots of (left) optimal two-photon Rabi frequencies and (right) optimal pulse durations as functions of Bragg order. These optimal pulse parameters give the maximum fidelity achievable for an $n$th order Bragg mirror process and a cloud of momentum width $\sigma$ (cf. figure~\ref{Fid_plots}). Lines joining the points are simply to guide the eye.}  
\end{figure}
Figure~\ref{Fid_plots} displays plots of the maximized fidelity $F_\pi(\Omega_\mathrm{opt},\tau_\mathrm{opt})$ as a function of the momentum width $\sigma$ for a variety of Bragg orders $n$. The first point to note is that for a fixed $n$, $F_\pi$ monotonically decreases with increasing $\sigma$. Thus the size of the cloud's momentum width places a fundamental limitation on the maximum transfer efficiency of the mirror. Secondly, for fixed $\sigma$, increasing $n$ does not necessarily change the maximum $F_\pi$ obtainable. A clear example of this is at $\sigma = 0.1 \,\hbar k$, where the maximum transfer efficiency is roughly $F_\pi \sim 0.9$ for $n \geq 2$. Furthermore, as shown in figure~\ref{Fid_plots}(b), $F_\pi(\Omega_\mathrm{opt},\tau_\mathrm{opt})$ is almost entirely independent of $n$ for higher order Bragg processes. The caveat, however, is that for larger $n$, maximal $F_\pi$ requires a larger $\Omega$, and hence a larger laser intensity (or smaller detuning). Our simulations show that, irrespective of $\sigma$, $\Omega_\mathrm{opt}$ scales roughly quadratically with $n$ (see figure~\ref{Mirror_opt_plots}). That is, optimal reflection occurs when $\Omega$ is on the order of the kinetic energy, $n^2 \omega_r$. This is consistent with the scaling in the analytic limit of the two-level Bragg regime of $\Omega \sim \left( (8 \, \omega_r)^{n-1}\left[ (n-1)!\right]^2\right)^{1/n}$ [cf. section~\ref{Sec_2_level}, and in particular (\ref{Omega_eff_2_level})]. As an aside, the typical values for $\tau_\mathrm{opt}$ ranged from $0.08-0.5 \, \omega_r^{-1}$ (see figure~\ref{Mirror_opt_plots}).  

As a simple application, we compared our calculations of the mirror efficiencies to measured values in the literature. In M\"uller \emph{et al.} \cite{Muller:2008b} mirror efficiencies between $80-90\%$ were reported for Bragg orders $n \leq 9$ with a source of momentum width $\sigma \sim 0.13 \, \hbar k$. This is consistent with our theory's prediction of $F_\pi \sim 0.85-0.9$, and also suggests that M\"uller \emph{et al.} were operating close to the fundamental mirror efficiency limit. Similarly, Chiow \emph{et al.} \cite{Chiow:2011} observed $6 \, \hbar k$ reflections of a BEC with efficiency $94\%$. From \cite{Anderson:1999}, we estimate the momentum width of their expanded cloud to be $\sigma \sim 0.05 \, \hbar k$, which places an upper bound of approximately $95\%$ on the mirror efficiency. In both cases, significant improvements to the mirror efficiency would only be possible by a further narrowing of the source momentum width.

Another stark feature seen in figure~\ref{Fid_plots}(a) is the `cusps', where at a particular $\sigma$ the plot undergoes a sudden change in slope. This $\sigma_\mathrm{cusp}$ corresponds to a change in the optimal operating regime. At $\sigma<\sigma_\mathrm{cusp}$ the atomic population is almost entirely divided between momentum distributions around the initial and target states, $\ket{-n \hbar k}$ and $\ket{n \hbar k}$ respectively. For $\sigma > \sigma_\mathrm{cusp}$ the optimal pulse results in non-negligible loss to other diffraction orders. 

There has been some interest in constructing LMT atom-optical elements with a series of lower order Bragg transitions, rather than a single higher order transition \cite{Chiow:2011}. Our results suggest that, provided one has ample laser power, a single $n$th order Bragg pulse will be more efficient than an $m$th order pulse applied $q$ times, where $q m = n$. For instance, for an atomic cloud with momentum width $0.2 \, \hbar k$ an $n = 10$ Bragg mirror process has a maximum fidelity of approximately 0.77. Thus about 23\% of the atoms will not be reflected, and in the simplest case would leave the interferometer. In contrast, a series of five $n=2$ pulses, where each pulse also has a maximum fidelity of 0.77, would have an effective maximum fidelity on the order of  $0.77^5 \approx 0.27$. Ignoring reductions in contrast, this loss of atom flux would reduce the SNR by a factor of two. Similar losses in atom flux would affect multiple Bragg pulse beamsplitters. This degradation in mirror and beamsplitter fidelity could be somewhat ameliorated by using an extremely narrow source, such as an atom laser. Nevertheless, all other considerations being equal, a single Bragg pulse will give a more efficient $2n \hbar k$ reflection than a series of lower order Bragg pulses.

\subsection{Two-level model for Bragg mirror} \label{Sec_2_level}
We can qualitatively understand the dependence of the fidelity on the momentum width $\sigma$, Bragg order $n$ and two-photon Rabi frequency $\Omega$ by considering the following simple model. Consider Bragg diffraction from a square pulse in the Bragg regime, where requirements of energy conservation are so severe that only two states are involved in the diffraction. More formally, the intermediate states between $\ket{(-n+\delta) \hbar k}$ and $\ket{(n+\delta) \hbar k}$ are adiabatically eliminated, and so the set of equations (\ref{c_eq}) reduce to \cite{Muller:2008}
\numparts
\begin{eqnarray} 
	i \dot{c}_{-n}(\delta, t) &= -2\omega_r n\delta c_{n}(\delta, t) + \frac{\Omega_\mathrm{eff}}{2}c_{n}(\delta,t) \label{Bragg_eqs_a} \\
	i \dot{c}_{n}(\delta, t) 	&=  2 \omega_r n\delta c_{n}(\delta, t) + \frac{\Omega_\mathrm{eff}}{2}c_{-n}(\delta,t), \label{Bragg_eqs_b}
\end{eqnarray}
\endnumparts
where $\Omega_\mathrm{eff}$ is the effective (or $2n$-photon) Rabi frequency. In the regime $\sigma \ll \hbar k$, $\Omega_\mathrm{eff}$ is approximately independent of $\delta$ and is given by
\begin{equation}
	\Omega_\mathrm{eff} = \frac{\Omega^n}{(8\, \omega_r)^{n-1}}\frac{1}{\left[(n-1)!\right]^2}. \label{Omega_eff_2_level}
\end{equation}
Solving (\ref{Bragg_eqs_a}) and (\ref{Bragg_eqs_b}), with initial condition $c_{-n}(\delta,0) = 1$ and $c_{n}(\delta,0) = 0$, for the target state population gives
\begin{equation}
	|c_{n}(\delta,t)|^2 = \frac{\Omega_\mathrm{eff}^2}{16 \omega_r^2 n^2 \delta^2+\Omega_\mathrm{eff}^2}\sin^2\left( \frac{t}{2}\sqrt{16 \omega_r^2 n^2 \delta^2+\Omega_\mathrm{eff}^2} \right). \label{2_level_pop}
\end{equation}
Comparing with the standard two-level atom, it is obvious that $4 \omega_r n \delta $ behaves precisely as a detuning from resonance. Given that most of the cloud is centred at $\delta = 0$, and we can get the maximum population transfer on resonance, we choose a pulse time $t = \pi/\Omega_\mathrm{eff}$.

\begin{figure}[t!]
\centering
\includegraphics[scale=0.31]{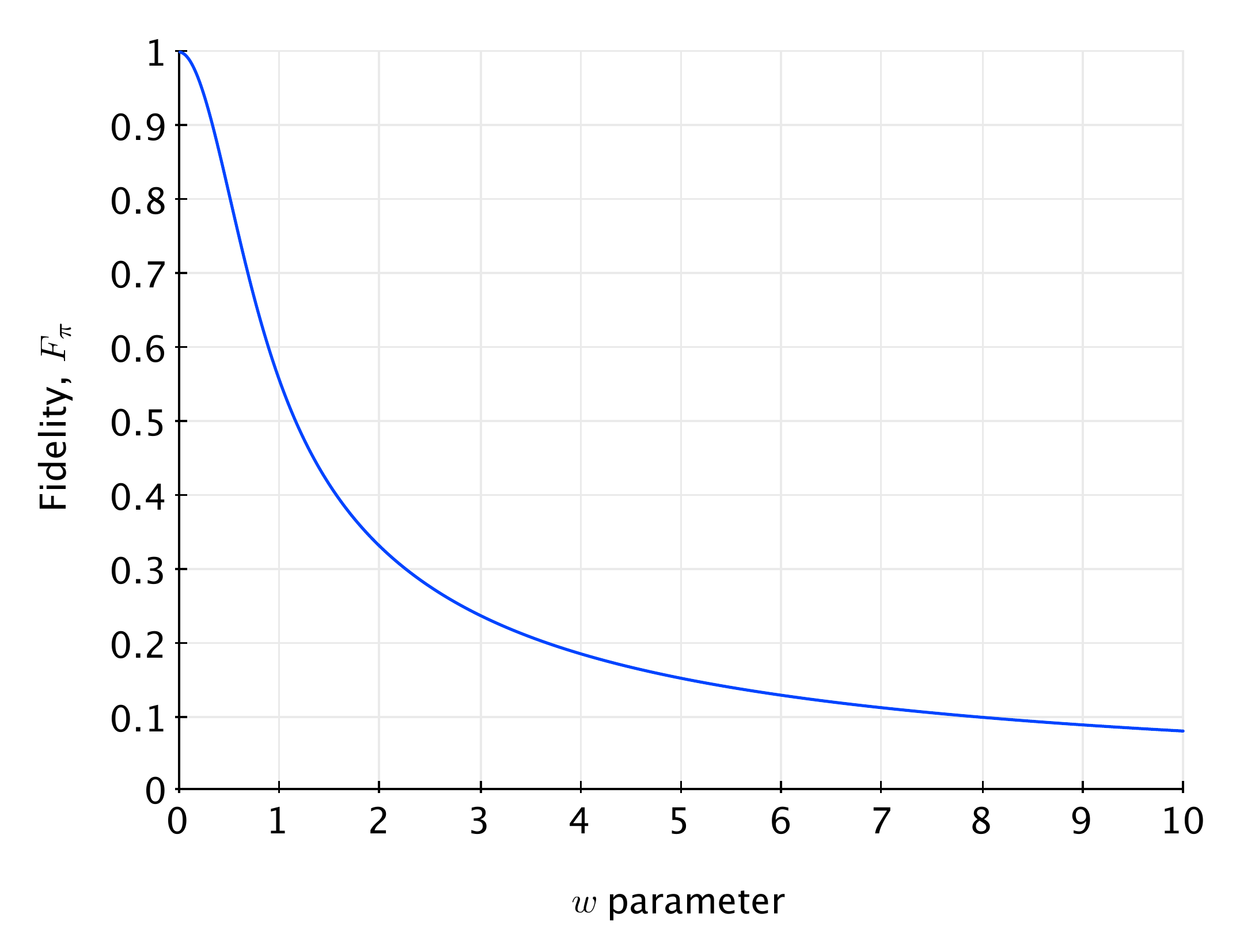}
\caption{\label{2_level_Fid_plot} Plot of the efficiency criterion as a function of the $w$ parameter [cf. (\ref{dim_F_pi})]. Note that the function monotonically decreases with increasing $w$.}
\end{figure}
Substituting (\ref{2_level_pop}) into the efficiency criterion (\ref{F_pi_criterion}) gives
\begin{equation}
	F_{\pi} = \int \frac{e^{-x^2/2w^2}}{\sqrt{2\pi w^2}}\frac{\sin^2\left(\frac{\pi}{2}\sqrt{1+x^2}\right)}{1+x^2} \, dx, \label{dim_F_pi}
\end{equation}
where we have transformed to the dimensionless variables
\begin{equation}
	x = \frac{4 \, \omega_r n \delta}{\Omega_{\mathrm{eff}}}; \qquad \qquad w = \frac{4 \, \omega_r n \sigma}{\Omega_{\mathrm{eff}}}. \label{rescaled_units}
\end{equation}
We can now see that in this two-level regime the only important parameter governing the efficiency of a Bragg process is $w$. A plot of $F_\pi$ as a function of $w$ is shown in figure~\ref{2_level_Fid_plot}. Noting that $F_\pi$ monotonically decreases with $w$, it is clear that the two-level model of the Bragg mirror predicts the following:
\begin{itemize}
	\item For a fixed $\Omega_\mathrm{eff}$ and $n$, an increase in momentum width $\sigma$ increases $w$, and thus decreases the efficiency of the process.
	\item For a fixed $\sigma$ and $\Omega_\mathrm{eff}$, an increase in the order of the Bragg process $n$ gives lower efficiency mirrors.
	\item If it is possible to increase $\Omega_\mathrm{eff}$, then decreases in efficiency due to increases in $n$ can be nullified. That is, for higher order Bragg processes and larger momentum width clouds, we can achieve a low $w$ (and hence a high $F_\pi$) by increasing the laser power.
\end{itemize}
This simple model qualitatively captures key features of our comprehensive numerical simulation of the Bragg mirror (cf. section~\ref{Mirror_numerics}). However, it does not \emph{quantitatively} give the correct maximum fidelities, nor the correct optimal pulse amplitudes and durations. Furthermore, the model predicts that we can compensate for increases in $\sigma$ by increases in $\Omega_\mathrm{eff}$. Figure~\ref{Fid_plots} shows that this is manifestly false for a realistic Bragg reflection. These disagreements are not surprising, given that the model only has validity in the regime where (a) the momentum width is much narrower than $\hbar k$; and (b) for each $\delta$, the intermediate states between $\ket{(-n+\delta) \hbar k}$ and $\ket{(n+\delta) \hbar k}$ are far detuned from the Bragg resonance (this bounds the effective Rabi frequency to $\Omega_\mathrm{eff} \ll 8\, \omega_r(n-1)^n/[(n-1)!]^2$). 

\subsection{Optimization with limitations on the two-photon Rabi frequency} \label{mirror_laser_limit}
The numerical results of figure~\ref{Fid_plots} show that the maximum fidelity remains virtually unchanged as one increases the Bragg order, on the proviso that one increases the two-photon Rabi frequency such that it is on the order of $\Omega \sim n^2\omega_r $. However, there will always be practical limitations on laser power and the loss that can be tolerated due to spontaneous emission. Both these constraints will place a clamp on the two-photon Rabi frequency. The two-level model from section~\ref{Sec_2_level} predicts that if there is a clamp on $\Omega_\mathrm{eff}$, then increasing the Bragg order $n$ past a certain point results in an increase in $w$, and hence a rapid decrease in the mirror fidelity. A comprehensive numerical analysis shows this result to be true in general, and furthermore shows that a clamp on the two-photon Rabi frequency limits high Bragg order, high fidelity reflections to narrow momentum width sources. 

\begin{figure}[t!]
\centering
\includegraphics[scale=0.45]{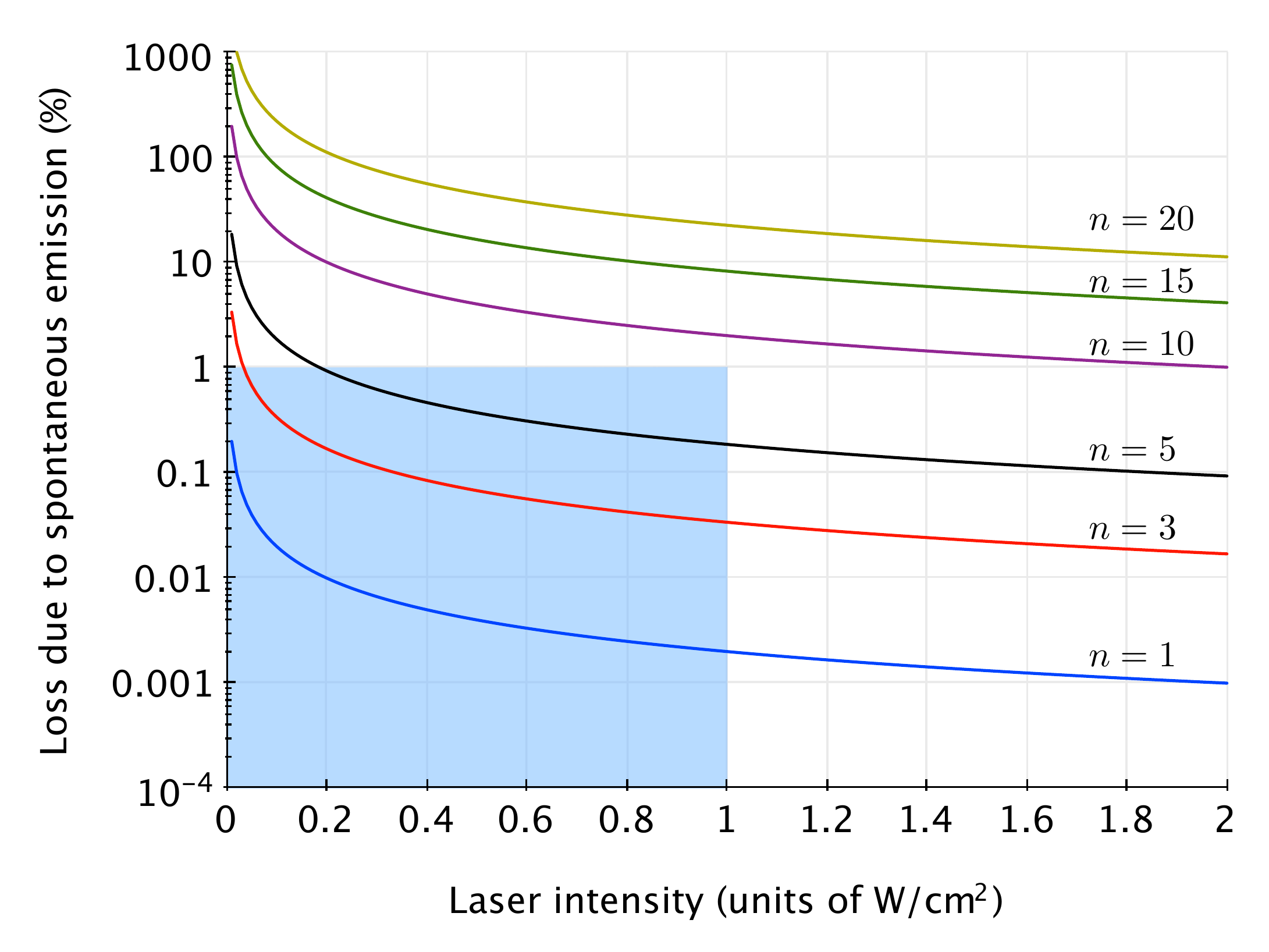}
\caption{\label{Sp_Em_plot} Plots of the percentage of atoms lost due to spontaneous emission as a function of the laser intensity for a range of Bragg orders. The curves shown have been calculated using the optimal pulse parameters shown in figure~\ref{Mirror_opt_plots} and are for a $\sigma = 0.1 \, \hbar k$ cloud; there is little change for other momentum widths less than $1 \, \hbar k$. The shaded rectangle indicates at what Bragg orders the maximum fidelity can be achieved, assuming that the interferometer cannot tolerate greater than $1 \%$ loss due to spontaneous emission, and that the maximum laser intensity is 1 W/cm$^2$.}
\end{figure}

We can estimate the Bragg orders where the maximum mirror fidelities shown in figure~\ref{Fid_plots} can be accessed. The spontaneous scattering rate is related to the intensity $I(t)$ and detuning $\Delta$ through the expression
\begin{equation}
	R\left(I(t), \Delta \right) = \frac{\Gamma}{4\pi}\frac{2 I(t)/I_\mathrm{sat}}{1+2 I(t)/I_\mathrm{sat}+4\Delta^2/\Gamma^2},
\end{equation}
where $\Gamma$ is the natural linewidth of the transition, $I_ \mathrm{sat}$ is the saturation intensity and $I(t) = I \exp(-t^2/2\tau^2)$, since the time-dependence of the intensity is exactly that of the two-photon Rabi frequency. Note that the Bragg mirror is created from two counter-propagating lasers of intensity $I$, and hence the spontaneous scattering rate has an effective intensity of $2 I$. The detuning can be written in terms of the two-photon Rabi frequency and intensity via the relations
\begin{equation}
	\Omega = \frac{\Omega_0^2}{2 \Delta} = \frac{(d E/\hbar)^2}{2\Delta} = \frac{d^2}{\hbar^2 \varepsilon_0 c}\frac{I}{\Delta},
\end{equation} 
where $d$ is the electric dipole moment of the atom, $c$ the speed of light in vacuum and $\varepsilon_0$ the permittivity of free space. Thus the proportion of atoms lost due to spontaneous emission is
\begin{eqnarray}
	S(\Omega, \tau, I) 	&= \int R\left(I(t), \Omega, \tau \right) dt \nonumber\\ 
					&= \frac{\Gamma}{4 \pi}\int \frac{2 I e^{-t^2/2\tau^2} /I_\mathrm{sat}}{1+2 I e^{-t^2/2\tau^2}/I_\mathrm{sat}+ 4 (d^2/ \hbar^2 \varepsilon_0 c)^2 (I^2/\Gamma^2 \Omega^2)} dt. \label{Sp_em}
\end{eqnarray}
By substituting the optimal pulse parameters (cf. figure~\ref{Mirror_opt_plots}) into equation (\ref{Sp_em}), we can calculate the loss due to spontaneous emission as a function of intensity for any order maximum fidelity $\pi$ pulse and source momentum width. As an example, consider a state-of-the-art laser system (such as that used in \cite{Muller:2008b}) that addresses the 780 nm D2 transition of a source of $^{87}$Rb atoms. This transition has a natural linewidth of $\Gamma = 2\pi \times 6.07$ MHz, an electric dipole moment $d = 2 \times 10^{-29}$ C/m, and a saturation intensity of $I_\mathrm{sat} = 1.6$ mW/cm$^2$. These numbers have been used to calculate the curves in figure~\ref{Sp_Em_plot}, which shows $S(\Omega_ \mathrm{opt}, \tau_ \mathrm{opt}, I)$ for a range of Bragg orders. Note that a loss over $100 \%$ indicates that all the atoms in the cloud have spontaneously scattered a photon before the pulse has finished. 

Now suppose that our experiment can tolerate no more than $1 \%$ of the atoms lost due to spontaneous emission, and furthermore we are limited to a laser intensity of 1 W/cm$^2$. Then the shaded region in figure~\ref{Sp_Em_plot} indicates the Bragg orders in our example scenario where the maximum mirror efficiency can be achieved. Specifically, maximum efficiency Bragg mirrors can certainly be achieved for $n \leq 5$; however they are unobtainable for $n \geq 10$. A maximally efficient 10th order Bragg mirror could be achieved by doubling the maximum available laser intensity to 2 mW/cm$^2$.

\begin{figure}[t!]
\centering
\includegraphics[scale=0.45]{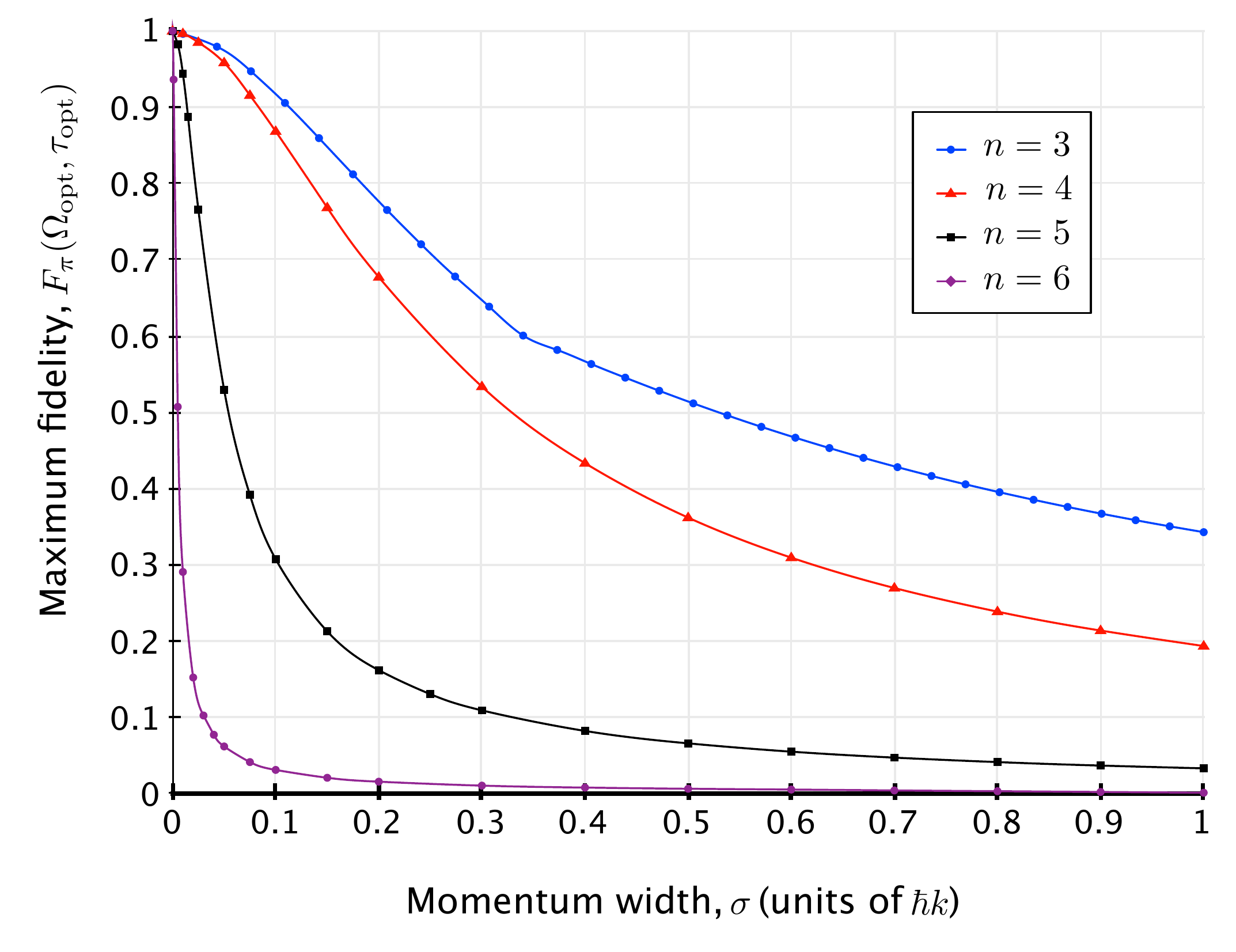}
\caption{\label{Fid_plot_Rabi_clamp} Plots of the maximum fidelity as a function of momentum width for a range of Bragg orders. The optimum two-photon Rabi frequency was subject to the constraint $\Omega_\mathrm{opt} \leq \Omega_\mathrm{max}$, where $\Omega_\mathrm{max} = 20 \, \omega_r$. For $n\geq 4$, the optimum $\Omega$ that gave the maximum fidelity was always $\Omega_\mathrm{opt} = \Omega_\mathrm{max}$. As before, the numerical data is represented by the points, and the lines are simply to guide the eye.} 
\end{figure}
Figure~\ref{Sp_Em_plot} illustrates that even with a state-of-the-art apparatus, maximum fidelity mirrors are practically unobtainable for moderate Bragg orders. This begs the question: how does a clamp on the two-photon Rabi frequency effect the maximum mirror fidelity? Towards answering this question, we numerically calculated some maximum fidelities for a two-photon Rabi frequency clamp of $\Omega_\mathrm{max} = 20 \, \omega_r$. The results of this investigation are shown in figure~\ref{Fid_plot_Rabi_clamp}. At $n = 3$, the optimal two-photon Rabi frequency $\Omega_\mathrm{opt}$ is always less than $\Omega_\mathrm{max}$, and so the clamp has no effect on the maximum fidelity. However, for $n \geq 4$, $\Omega_\mathrm{max}$ is less than the optimal two-photon Rabi frequency. This constrains the allowable optimum to $\Omega_\mathrm{opt} \leq \Omega_\mathrm{max}$, the effect of which is to drastically reduce the fidelity. In particular, figure~\ref{Fid_plot_Rabi_clamp} illustrates two points. Firstly, when the two-photon Rabi frequency is constrained increases in $n$ result in dramatic decreases in the maximum obtainable fidelity. The second point to note is that larger Bragg orders require an increasingly narrow momentum width in order to obtain good transfer efficiencies. For example, at $n=3$ a fidelity greater than 0.9 can be achieved for any $\sigma \lesssim 0.1 \, \hbar k$. For $n=5$ and $n=6$, the momentum width needs to be $\sigma \lesssim 0.015\, \hbar k$ and $\sigma \lesssim 0.001 \, \hbar k$, respectively, in order to achieve a Bragg mirror with fidelity greater than $0.9$. 

Clamping the the two-photon Rabi frequency changes both the optimal two-photon Rabi frequency and the optimal pulse duration. When $\Omega_\mathrm{max} = 20 \, \omega_r$, the optimal pulse time increases from $\tau_\mathrm{opt} \sim 0.4-0.7 \, \omega_r^{-1}$ for $n=4$ to $\tau_\mathrm{opt} \sim 2-2.7 \, \omega_r^{-1}$ for $n=5$ and $\tau_\mathrm{opt} \sim 19-26 \, \omega_r^{-1}$ for $n=6$. For the $^{87}$Rb 780 nm D2 transition, this translates to pulse durations on the order of 1-10 ms. For pulse times considerably longer or shorter than these optimal values, the fidelity will be close to zero.

To summarize, limits on laser power and spontaneous loss place an upper bound on the two-photon Rabi frequency. Clamping the two-photon Rabi frequency reduces the efficiency of higher order Bragg processes by a factor that strongly depends upon the momentum width of the cloud. As $n$ increases, this degradation in fidelity becomes more severe and can only be overcome by increasingly narrow momentum widths and longer pulse durations.

\begin{figure}[t!]
\centering
\includegraphics[scale=0.37]{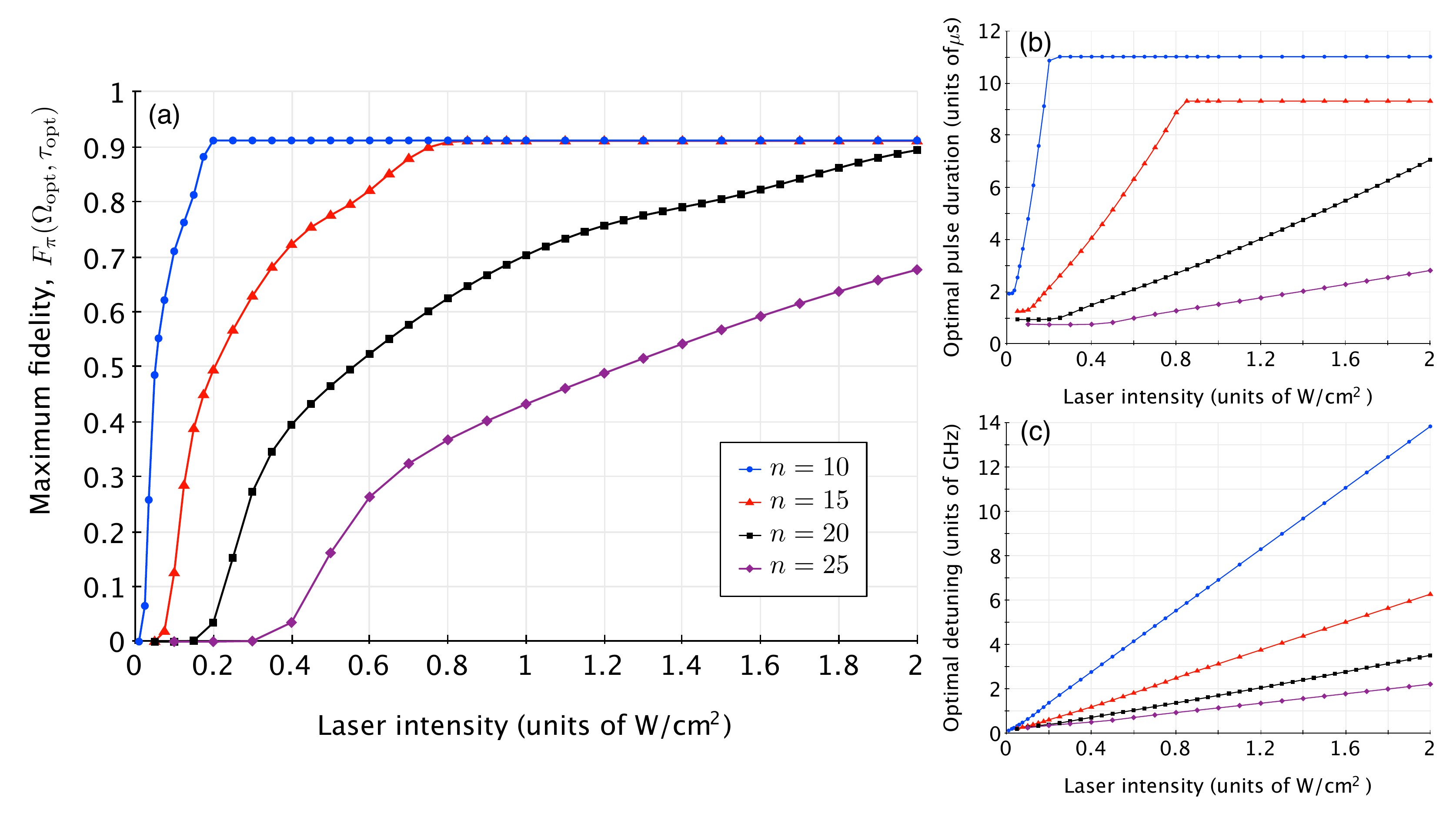}
\caption{\label{clamped_SpEm_plots} (a) Plots of the maximum fidelity as a function of laser intensity for a $\sigma = 0.1 \hbar k$ cloud, where the two-photon Rabi frequency has been constrained by equation (\ref{Omega_max_int}). For each Bragg order, the optimal pulse duration and detuning are given as functions of intensity in (b) and (c) respectively. The lines are simply to guide the eye.}
\end{figure}

We now calculate the maximum fidelities that can practically be achieved for the experimental scenario considered when calculating figure~\ref{Sp_Em_plot}. For sufficiently large detuning, we can ignore the time-dependence in the denominator of equation (\ref{Sp_em}). In this case
\begin{equation}
	S \approx  \frac{\sqrt{2\pi}\Gamma \tau}{4 \pi} \frac{2 I /I_\mathrm{sat}}{1+2 I /I_\mathrm{sat}+ 4 (d^2/ \hbar^2 \varepsilon_0 c)^2 (I^2/\Gamma^2 \Omega^2)}.
\end{equation} 
Rearranging this expression gives, for a fixed $S$, the maximum two-photon Rabi frequency as a function of $I$ and $\tau$:
\begin{equation}
	\Omega_\mathrm{max}(I,\tau) = \frac{(2 d^2/\hbar^2 c \varepsilon_0 \Gamma) I}{\sqrt{\left\{ \Gamma \tau/(\sqrt{2\pi} S) - 2\right\}(I/I_\mathrm{sat}) -1}}. \label{Omega_max_int}
\end{equation} 
Figure~\ref{clamped_SpEm_plots}(a) shows the effect of this constraint on the maximum fidelities for $S = 0.01$ and a cloud of momentum width $\sigma = 0.1 \, \hbar k$. It is clear from this plot that a maximally efficient Bragg reflection with $F_\pi \approx 0.92$ (cf. figure~\ref{Fid_plots}) requires some minimum laser intensity $I_\mathrm{min}$, and that this minimum intensity increases with increasing Bragg order. For intensities less than $I_\mathrm{min}$, the efficiency of the Bragg mirror is vastly reduced. Figures~\ref{clamped_SpEm_plots}(b) and (c) show the optimal pulse durations and detunings, respectively, needed to achieve the constrained maximum fidelities of figure~\ref{clamped_SpEm_plots}(a).

\section{Analysis for Mach-Zehnder interferometer}
In section~\ref{Bragg_results}, we used loss from the target state as a characterization of Bragg mirror efficiency. However, the phase of the atoms also has an important effect on the SNR. We therefore consider the effect of losses and interference by extending our analysis to a MZ interferometer. We show that the MZ interferometer shares many of the trends exhibited by the Bragg mirror, including limitations on the SNR due to the momentum width. Furthermore, we calculate the optimal Bragg order and (for a thermal source) momentum width that maximize the SNR. Finally, we compare the SNR for optimized thermal and Bose-condensed sources. 

\subsection{Efficiency criterion for Mach-Zehnder interferometer} \label{Efficiency_MZ}
For a MZ interferometer, we ultimately want to maximize the SNR (\ref{SNR}) for a phase measurement. A good figure of merit will be proportional to the SNR, and will incorporate losses and phase information. Losses are accounted for by simply counting the number of atoms in the two output states of the interferometer. Phase information is related to the SNR via the fringe contrast of the relative populations. We thus define the MZ interferometer efficiency criterion as
\begin{equation}
G \equiv \mathcal{C} \sqrt{N}.
\end{equation}
It is not difficult to generalize the contrast $\mathcal{C}$ and the atomic population $N$ in the two output ports of the interferometer to the case of finite momentum width atomic clouds. Define the \emph{total population of momentum state} $m \hbar k$ as
\begin{equation}
	P_m(\phi) = \int \frac{1}{\sqrt{2\pi \sigma^2}} e^{-\delta^2/ 2 \sigma^2} \left|c_{m}(\delta,\phi,t_f)\right|^2 \, d\delta, \label{P_m}
\end{equation}
where $t_f$ is now the time after all pulses have interacted with the atomic cloud. Without loss of generality, we normalize the total number of atoms to 1. In this case, the coefficients $c_m(\delta,\phi,t)$ are calculated from the set of equations (\ref{c_eq}) with pulse sequence (\ref{3_pulse}). The relative total population in state $n \hbar k$ is then
\begin{equation}
	P^\mathrm{rel}_n(\phi) = \frac{P_n(\phi)}{P_{-n}(\phi)+P_n(\phi)}.
\end{equation}
The contrast is given as the difference between maximum and minimum relative total populations:
\begin{equation}
	\mathcal{C} = \underset{0\leq\phi\leq 2\pi/n}{\max}P^\mathrm{rel}_n(\phi)-\underset{0\leq\phi\leq 2\pi/n}{\min}P^\mathrm{rel}_n(\phi).
\end{equation}
Note that for an $n$th order Bragg process, the phase is imprinted on the atoms $n$ times. Hence the period of any fringes will be $2\pi/n$. If the fringes are approximately sinusoidal, which appears to be true even for moderately large momentum widths, then $\max_{\,0\leq\phi\leq 2\pi/n}P^\mathrm{rel}_n(\phi) = P^\mathrm{rel}_n(\pi/n)$ and $\min_{\,0\leq\phi\leq 2\pi/n}P^\mathrm{rel}_n(\phi) = P^\mathrm{rel}_n(0)$. We assume the MZ interferometer operates at mid-fringe, and it is at this point that we determine the total number in the two output ports: $N = P_{-n}(\pi/2n)+P_n(\pi/2n)$. Therefore the $G$-factor is calculated via the expression
\begin{equation}
G \equiv \left( \underset{0\leq\phi\leq 2\pi/n}{\max}P^\mathrm{rel}_n(\phi)-\underset{0\leq\phi\leq 2\pi/n}{\min}P^\mathrm{rel}_n(\phi)\right)\sqrt{P_{-n}(\phi/2n)}. \label{G_factor}
\end{equation}

\subsection{Two-level model for the Mach-Zehnder interferometer} \label{2_level_MZ}
As in section~\ref{Sec_2_level}, we can gain qualitative understanding by considering the behaviour of an MZ interferometer composed of square pulses operating in the Bragg regime. The solution to (\ref{Bragg_eqs_a}) and (\ref{Bragg_eqs_b}) can be expressed as the unitary matrix
\begin{equation}
	U(\tilde t,\phi)=\begin{pmatrix}
		 	\cos \left( \frac{\sqrt{1+x^2}}{2}\tilde t\right) - i \frac{x}{\sqrt{1+x^2}}\sin\left(  \frac{\sqrt{1+x^2}}{2}\tilde t \right) &	\frac{-i \exp\left(-i n \phi\right)}{\sqrt{1+x^2}}\sin\left(  \frac{\sqrt{1+x^2}}{2}\tilde t \right) \\
		\frac{-i\exp\left(i n \phi\right)}{\sqrt{1+x^2}}\sin\left(  \frac{\sqrt{1+x^2}}{2}\tilde t \right)	&	\cos \left( \frac{\sqrt{1+x^2}}{2}\tilde t \right) + i \frac{x}{\sqrt{1+x^2}}\sin\left(  \frac{\sqrt{1+x^2}}{2}\tilde t \right)
	\end{pmatrix},
\end{equation}
where we use the dimensionless units $x = 4 \omega_r n \delta /\Omega_\mathrm{eff}$ and $\tilde t = \Omega_\mathrm{eff} t$. The phase factor $\exp(\pm i n \phi)$ is due to the laser pulse phase, $\phi$. Formally, this is included by the replacement $\Omega_\mathrm{eff} \to \Omega_\mathrm{eff} \exp\left(i n \phi\right)$ in (\ref{Bragg_eqs_a}) and (\ref{Bragg_eqs_b}). For the above unitary matrix, $\pi/2$ and $\pi$ pulses are given by setting $\tilde t = \pi/2$ and $\tilde t = \pi$, respectively. 

\begin{figure}[t!]
\centering
\includegraphics[scale=0.3]{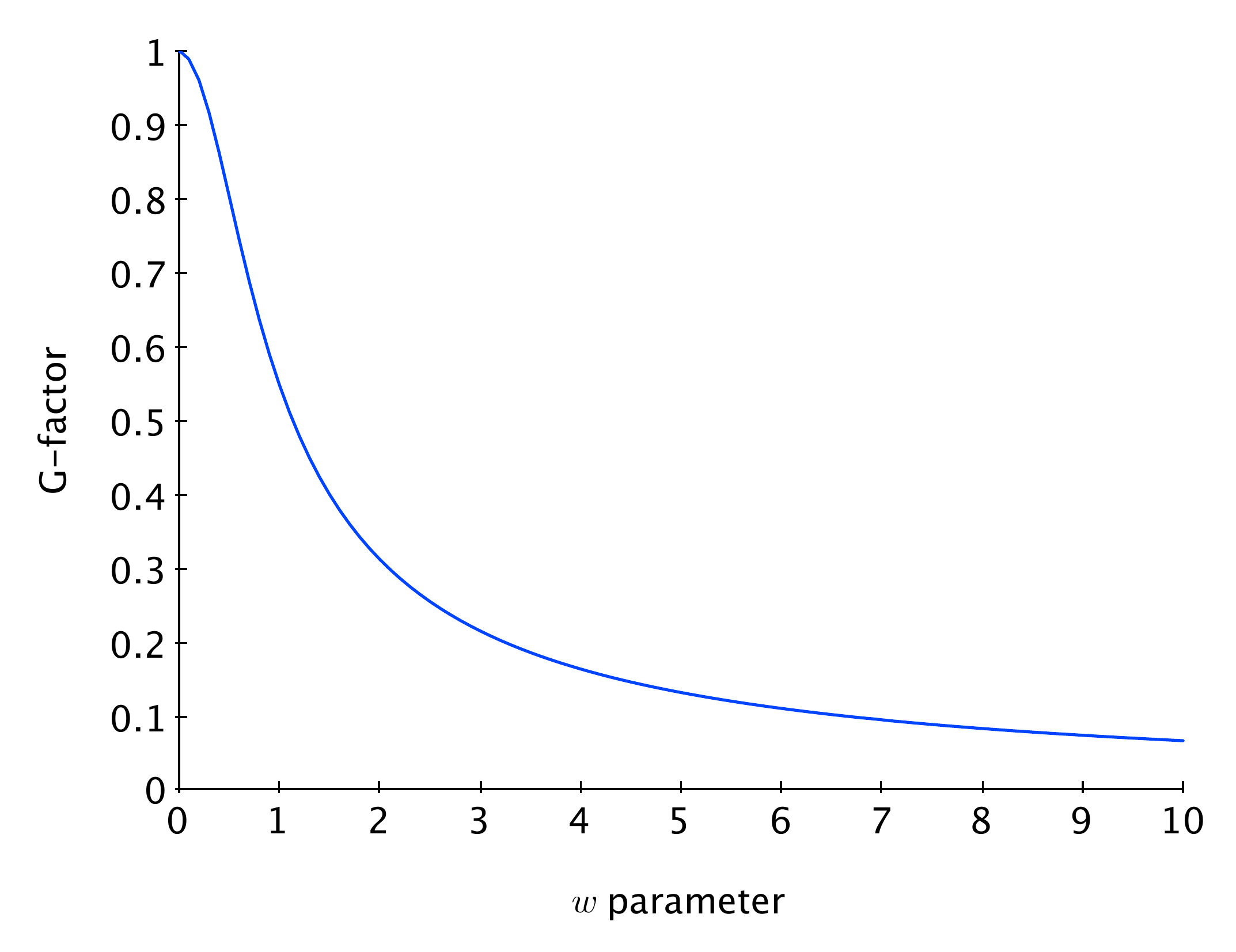}
\caption{\label{2_level_G_plot} Plot of the $G$-factor as a function of $w$. The key feature to note is that $G$ decreases with $w$ (i.e. decreases with increasing momentum width $\sigma$ or Bragg order $n$). However, one can decrease $w$ to obtain the same $G$-factor by increasing the effective Rabi frequency.}
\end{figure}

During the time between pulses, the atoms undergo free evolution as described by
\begin{equation}
	U_f(\tilde t) = 	\begin{pmatrix}
				\exp\left(i x \tilde t/2\right)	&	0 \\
				0		&	\exp\left(-i x \tilde t/2\right)
			\end{pmatrix}. 
\end{equation}
We can see then that the output of a MZ interferometer with three pulses, with an overall phase shift $\phi$ included on the third pulse, is given by
\begin{equation}
	\ket{\psi_\mathrm{output}} = U(\pi/2,\phi)U_f(T)U(\pi,0)U_f(T)U(\pi/2,0)\ket{\psi_\mathrm{input}}.
\end{equation}
Here $T$ is the interrogation time between pulses. We choose the initial state $\ket{\psi_\mathrm{input}}$ such that all the atoms are initially in the ground state. $\ket{\psi_\mathrm{output}}$ can be used to find the total population of the $-n\hbar k$ and $n \hbar k$ states via (\ref{P_m}). These can be used to calculate the $G$-factor (\ref{G_factor}). Provided $T$ is much larger than the pulse duration, $G$ will only depend on the $w$ parameter defined in (\ref{rescaled_units}). A plot of $G$ as a function of $w$ is shown in figure~\ref{2_level_G_plot}. The same conclusions that applied to the Bragg mirror also apply to $G$, and hence the SNR of the interferometer: (a) larger momentum widths and/or higher Bragg orders decrease $G$, and (b) decreases in $G$ due to increases in Bragg order can be neutralised by increasing the effective Rabi frequency. However the same limitations on the two-level model of the Bragg mirror also apply to this model for the MZ interferometer: it is only valid in the regime $\sigma \ll \hbar k$ and $\Omega_\mathrm{eff} \ll 8\, \omega_r(n-1)^n/[(n-1)!]^2$. 

\subsection{Numerical simulation for a Mach-Zehnder interferometer} \label{Results_MZ}
We performed a numerical optimization for a pulse sequence in the MZ configuration. This problem requires the optimization of beamsplitter and mirror pulses, parametrized by $(\Omega_\mathrm{bs}, \tau_\mathrm{bs})$ and $(\Omega_\mathrm{m}, \tau_\mathrm{m})$ respectively, such that the $G$-factor (\ref{G_factor}) is maximized. The optimal amplitude and duration of the mirror pulse was chosen based on the values calculated by the independent optimization of the Bragg mirror considered in section~\ref{Bragg_results}. The interrogation time $T$ was always chosen to be much larger than any of the pulse durations. In this regime, we expect $G$ to be approximately independent of $T$. Thus the optimization of the MZ interferometer reduces to an optimization of the amplitude and duration of the beamsplitter pulses.

\begin{figure}[t!]
\centering
\includegraphics[scale=0.44]{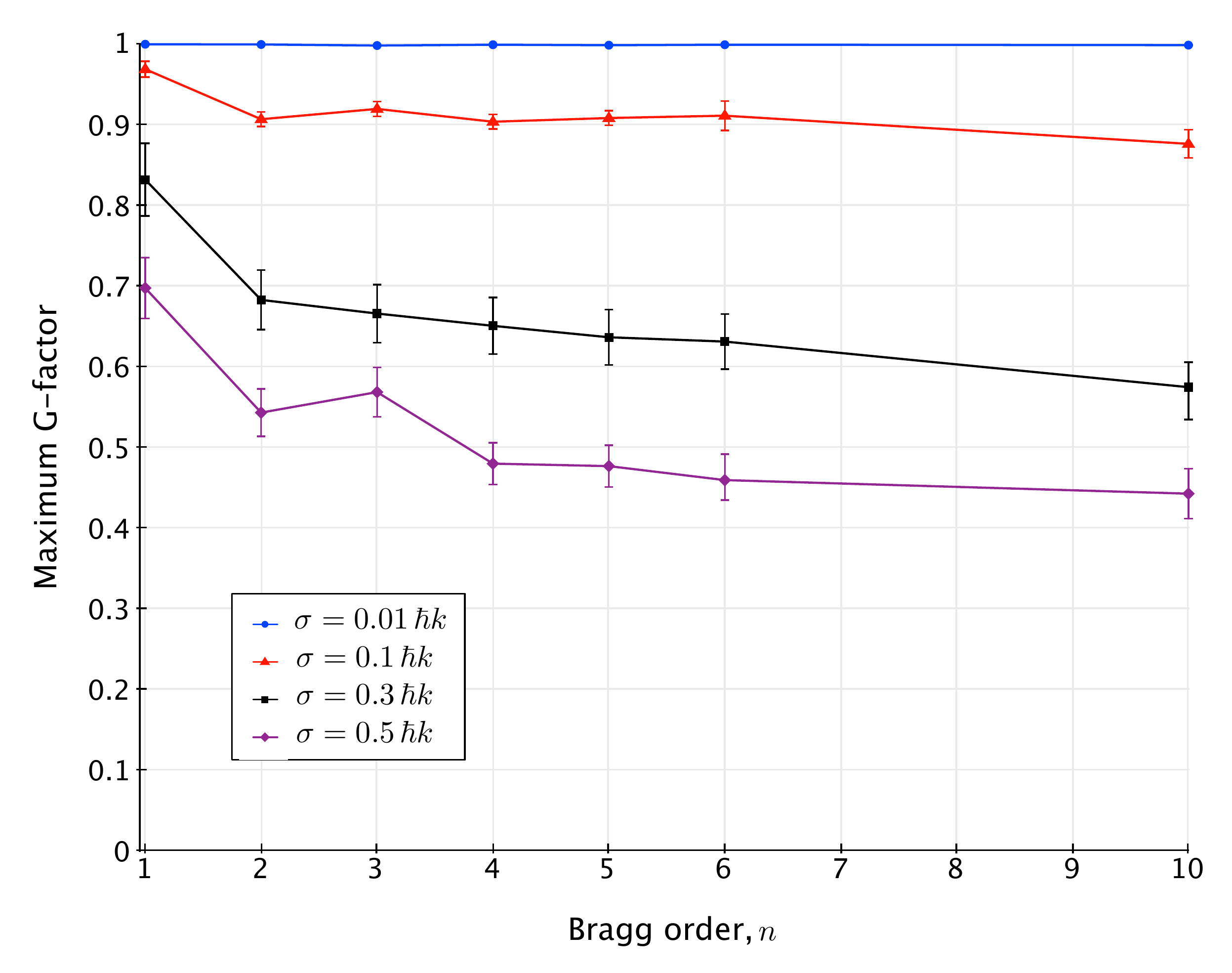}
\caption{\label{G_plot} Plots of the $G$-factor as a function of Bragg order, $n$, for different momentum widths. Recall that the total atom number has been normalized to 1. No constraint has been placed on the two-photon Rabi frequency. The error bars are estimates of the uncertainty in the maximum $G$-values. Lines joining points are just to guide the eye.}
\end{figure}

The numerical results for the optimization of $G$ as a function of $n$ and $\sigma$ with unconstrained $\Omega$ are shown in figure~\ref{G_plot}. As with the Bragg mirror, it is clear that increasing the momentum width leads to a reduction in the maximum $G$ obtainable. The other interesting feature is that, apart from the sharp drop in maximum $G$ for the lowest two or three $n$, the maximum $G$-factor is approximately constant with increasing Bragg order, at least for the Bragg orders and momentum widths shown here. Again the MZ interferometer behaves similarly to the Bragg mirror; with ample laser power the interferometer SNR does not appreciably decrease with increasing Bragg order. 

Whilst $G$ should be independent of the interrogation time $T$ in the limit $T \to \infty$, computational limitations sometimes necessitated that the simulations be run with an interrogation time somewhat smaller than the interrogation times used in typical experiments (although $T \gg \tau_\mathrm{m}, \tau_\mathrm{bs}$ was always satisfied). This meant that our numerical calculations of maximum $G$ had a slight $T$ dependence. Calculations of maximum $G$ for different values of $T$ allowed us to estimate the maximum $G$ in the limit $T \to \infty$, and place uncertainties on this estimate. The points and error bars in figure~\ref{G_plot} are an estimate of the maximum $G$-factor in the limit $T \to \infty$ and the uncertainty on that estimate, respectively. Maximum $G$ and associated uncertainties were calculated similarly for the remaining plots in this paper.

We noted in section~\ref{Mirror_numerics} the approximate quadratic scaling of the optimal mirror pulse amplitude with Bragg order. Similarly, the optimal beamsplitter pulse scales approximately quadratically with $n$. Furthermore, the optimized values of $(\Omega_\mathrm{bs}, \tau_\mathrm{bs})$ did, in general, differ to the optimal values of $(\Omega_\mathrm{m}, \tau_\mathrm{m})$, but by no more than a factor of two (and usually much less). 

\begin{figure}[t!]
\centering
\includegraphics[scale=0.55]{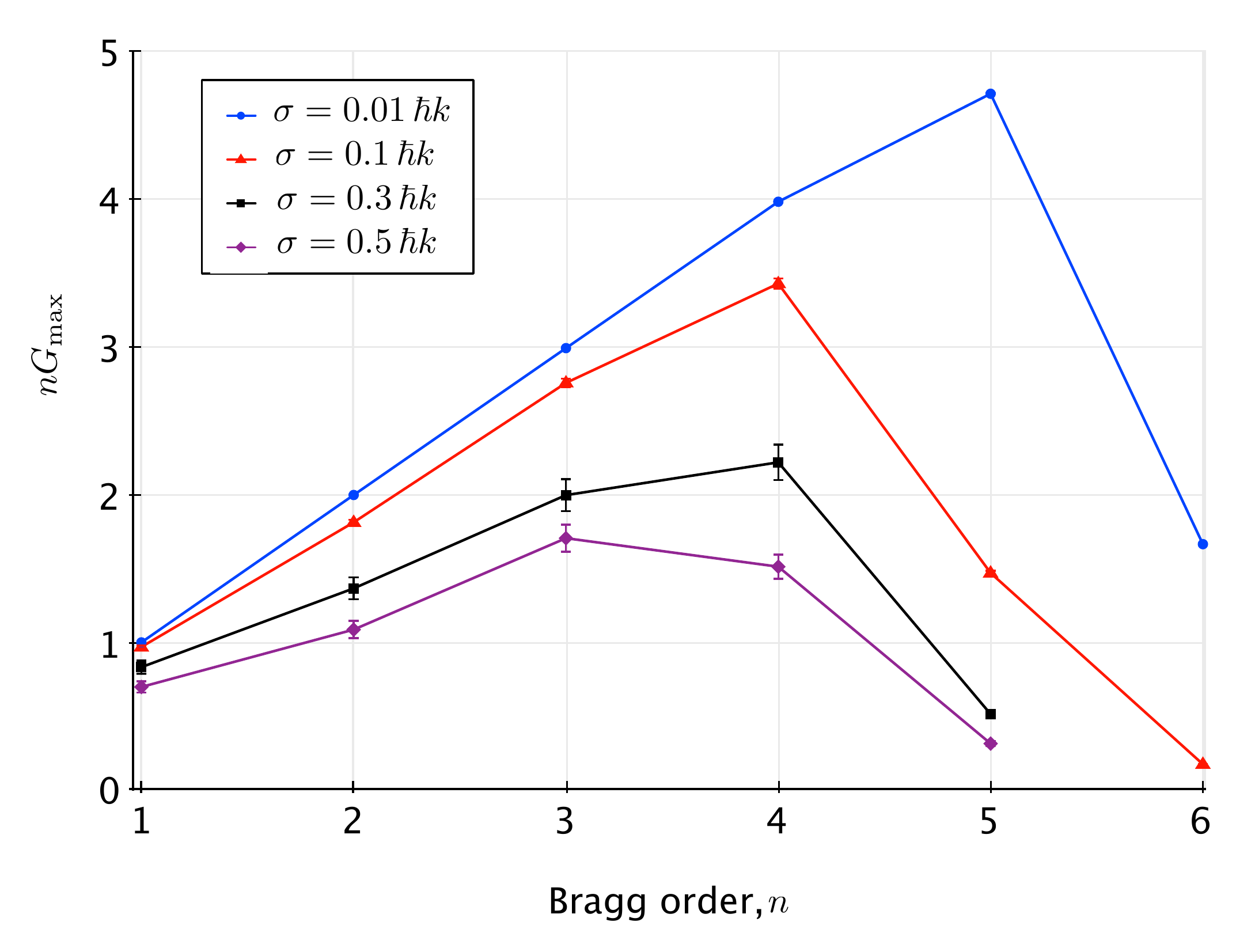}
\caption{\label{nG_plot_clamped} Plots of $nG_\mathrm{max}$ as a function of Bragg order, $n$, for different momentum widths. The two-photon Rabi frequency has been clamped at $\Omega_\mathrm{max} = 20 \, \omega_r$. Note that for each $\sigma$ there exists an optimal Bragg order $n$ that maximizes $n G_\mathrm{max}$ (and therefore the SNR). Lines joining points are just to guide the eye.}
\end{figure}

We also considered the effect of a two-photon Rabi frequency clamp on the $G$-factor. Clamping at $\Omega_\mathrm{max} = 20\,\omega_r$ resulted in significant decreases in the $G$-factor for $n \geq 4$. Whilst such decreases were ameliorated by a narrower momentum width, overall the maximum $G$ decreased for increasing $n$. More interestingly, we compared how a clamped two-photon Rabi frequency affected $n G_\mathrm{max}$ (which is proportional to the SNR), where $G_\mathrm{max}$ is the maximum $G$-factor for a particular $\sigma$ and $n$. As shown in figure~\ref{nG_plot_clamped}, for a fixed $\sigma$ there exists a Bragg order that maximizes $n G_\mathrm{max}$. Furthermore, this optimum Bragg order appears to decrease as $\sigma$ increases. This is a key result, for in practice there will \emph{always} exist some upper bound on the two-photon Rabi frequency. Consequently, it is simply not true that increasing the momentum transferred by the beamsplitters and mirrors will always improve the SNR, even at the shot noise limit.

\subsection{Comparison of thermal and Bose-condensed sources} \label{Sec_BEC_vs_thermal}
Na\"ively, one may think that figure~\ref{nG_plot_clamped} implies that a narrower momentum width gives a larger SNR for a MZ interferometer. This is only true if the atom flux is fixed for different momentum widths. In reality, broader sources tend to have a higher flux than narrower sources. One therefore expects there to be some trade-off between increasing the $G$-factor by narrowing the momentum width, and increasing the atom flux by broadening the momentum width. In this final section, we include the flux considerations due to choice of source in order to compare the SNR of thermal and Bose-condensed sources. 

We can incorporate the Bragg order and the effect of source type on flux into our analysis by defining the \emph{effective $G$-factor}:
\begin{equation}
	G_\mathrm{eff}	\equiv n \sqrt{N_i} G_\mathrm{max},
\end{equation}
where $N_i$ is the flux of the source that is input into the interferometer. $G_\mathrm{eff}$ is directly proportional to the SNR, and includes considerations due to the initial flux of the source, momentum transferred by the beamsplitters and mirrors, phase information (via the fringe contrast) and losses from the two interferometer modes during the interferometric process. 

Thermal sources are typically velocity-selected before the first beamsplitting pulse. Assuming a velocity-selection pulse reduces the cloud to a Gaussian distribution of width $\sigma$, then the fraction of atoms remaining after the velocity-selection pulse is proportional to $\sigma/\sigma_0$, where $\sigma_0$ is the initial momentum width of the thermal source. We therefore assume $N_i = \sigma/\sigma_0$ for a thermal source, where for convenience we have normalized the flux of the initial thermal source of width $\sigma_0$ to 1. By maximizing $G_\mathrm{eff}$ we can not only determine the optimal Bragg order, but also the optimal velocity-selection.

Unlike a velocity-selected thermal source, the input flux of Bose-condensed sources does not linearly depend upon the momentum width. BEC and atom laser fluxes typically depend upon a multitude of factors that are highly apparatus dependent. For the purposes of comparing $G_\mathrm{eff}$ for thermal and Bose-condensed sources, we assume the flux of an expanded BEC or atom laser to be a factor of 25 lower than a $1 \, \hbar k$ thermal source. This is the difference in flux between the state-of-the-art BEC machine in \cite{van_der_Stam:2007} and the best published state- and velocity-selected thermal cloud used in a gravimeter \cite{Muller:2008c}. We also assume an expanded BEC and atom laser  have momentum widths of $0.1 \, \hbar k$ \cite{Debs:2011} and $0.01 \, \hbar k$ \cite{Jeppesen:2008}, respectively. 

\begin{figure}[t!]
\centering
\includegraphics[width=\textwidth]{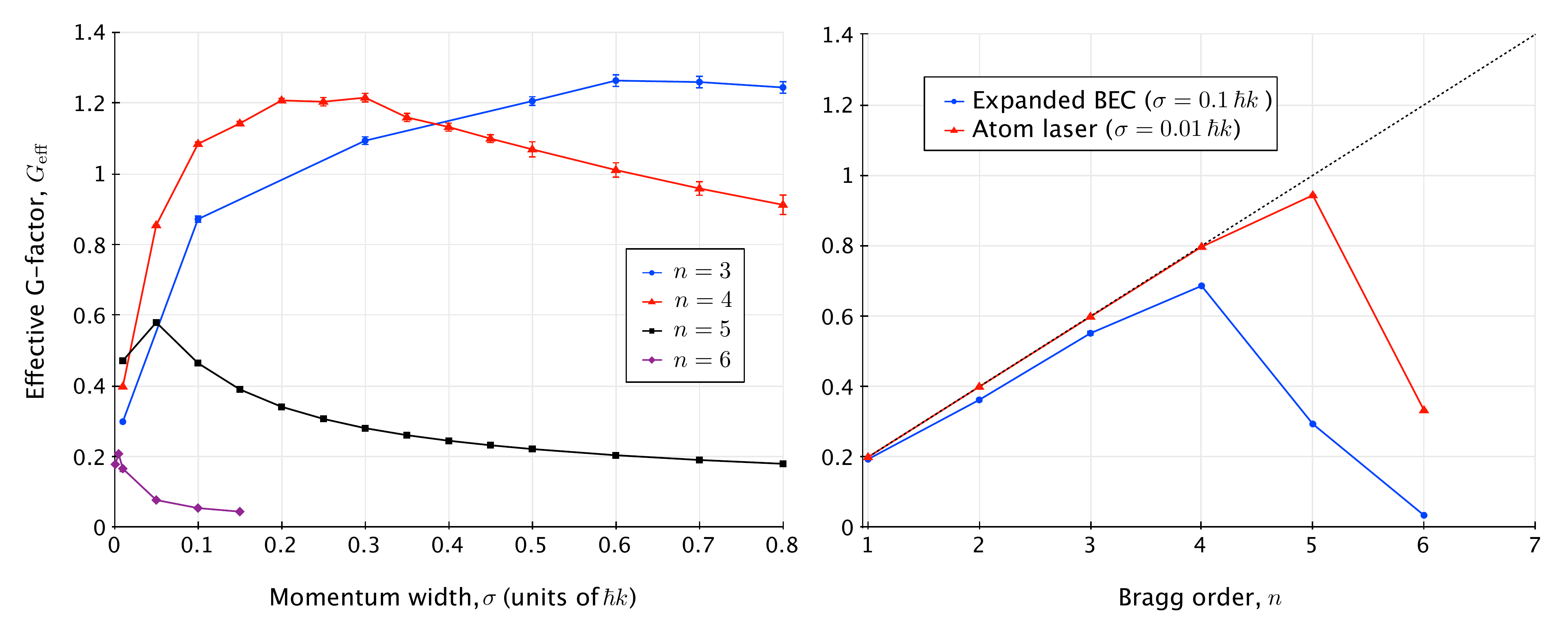}
\caption{\label{BEC_vs_thermal} Plots comparing the effective $G$-factor for thermal and BEC sources. The plot on the left-hand side displays, for different Bragg orders, $G_ \mathrm{eff}$ as a function of momentum width for a thermal source initially of width $\sigma_0 = 1 \, \hbar k$ velocity-selected to momentum width $\sigma$. The plot on the right-hand side shows $G_\mathrm{eff}$ as a function of Bragg order for an expanded BEC source of $\sigma = 0.1 \, \hbar k$ and an atom laser of $\sigma = 0.01 \, \hbar k$. The dotted line in the right-hand figure is $G_\mathrm{eff}$ in the limit of a plane-wave Bose-condensed source. We have assumed the Bose-condensed sources to have a factor of 25 less flux than the $1 \hbar k$ thermal source. The scale for $G_\mathrm{eff}$ is the same for the thermal and Bose-condensed sources, thus allowing for easy comparison. These data assume an upper bound on the two-photon Rabi frequency of $\Omega_\mathrm{max} = 20 \, \omega_r$.}
\end{figure}
Figure~\ref{BEC_vs_thermal} graphically compares the effective $G$-factor for thermal and Bose-condensed sources. For a thermal source, these data indicate that a better SNR is obtained by velocity-selecting less (giving a larger initial flux) and operating at a lower Bragg order where SNR reductions due to the broader momentum width and limited laser power are reduced. Indeed, the optimal pulse amplitudes and times for $n=3$ are unaffected by the $\Omega_\mathrm{max} = 20 \, \omega_r$ clamp we have considered, whilst the clamp drastically reduces $G_\mathrm{max}$ for the 5th and 6th Bragg orders. In contrast, the narrow momentum widths of Bose-condensed sources allow for low loss, high contrast operation at higher Bragg orders than any thermal source. For the state-of-the-art parameters consider here, the best SNR for an expanded BEC and atom laser occur at $n=4$ and $n=5$, respectively. The theoretical SNR of a plane-wave Bose-condensed source, also shown in figure~\ref{BEC_vs_thermal}, indicates that further narrowing of the momentum width of Bose-condensed sources allows for operation at increasingly larger Bragg orders. It appears, therefore, that LMT atom-optical elements are more beneficial for increasing the SNR for Bose-condensed sources than for thermal sources. 

From figure~\ref{BEC_vs_thermal}, we can see that current state-of-the art thermal and Bose-condensed sources theoretically give, within a factor of two, the same SNR for a MZ interferometer. This shows that with the application of LMT beamsplitters and mirrors, Bose-condensed sources are competitive with thermal sources, even with a factor of 25 less flux. Technological improvements that increase the flux of Bose-condensed sources are not inconceivable, and could be achieved by increasing the atom number of the condensate and/or increasing the repetition rate of the BEC machine. For the situation considered in figure~\ref{BEC_vs_thermal}, an atom laser with a comparable flux to a $1 \, \hbar k$ thermal source would give a SNR four times larger than the best SNR achievable with this thermal source. It is likely that Bose-condensed sources have further advantages when the momentum width transverse to the direction of the optical lattice is considered. Our analysis entirely ignored the transverse momentum width, and consequently effects on the SNR due to wavefront distortions and the Coriolis effect were neglected. Based on work by Debs \emph{et al.} \cite{Debs:2011} and Louchet-Chauvet \emph{et al.} \cite{Louchet-Chauvet:2011}, we expect narrower momentum width sources to be less susceptible to aberrations and the Coriolis effect. 

\section{Conclusions}
We have developed a relatively straightforward theoretical method for optimizing Bragg atom-optical elements. We have used this method to investigate the effect of momentum width on the efficiency of Bragg mirrors and Bragg MZ atom interferometers. Our comprehensive numerical analysis of the Bragg mirror (MZ interferometer), coupled with the qualitative understanding given by a two-level model, demonstrated that (a) the source momentum width \emph{fundamentally} limits the efficiency of Bragg mirrors (SNR of MZ interferometers); (b) for fixed momentum width, increasing the Bragg order does not generally change the maximum fidelity ($G$-factor) provided one has sufficient laser power; (c) the optimal two-photon Rabi frequency for mirror and beamsplitter pulses scales roughly quadratically with the Bragg order; and (d) a limitation on the two-photon Rabi frequency will, above some Bragg order, result in drastic decreases in the maximum mirror efficiency ($G$-factor), and further restricts a high fidelity ($G$-factor) to narrow momentum width clouds. Finally, we incorporated source type and, for a particular two-photon Rabi frequency clamp, calculated the optimal Bragg order and (for a thermal source) velocity selection that maximized the SNR. We compared thermal and Bose-condensed sources, and showed that state-of-the-art Bose-condensed sources are competitive with state-of-the-art thermal sources. 

The analysis undertaken in this paper shows that, fundamentally, momentum width \emph{does} matter for atom interferometry with Bragg pulses. Moreover, we have demonstrated that this analysis can be used to optimize the atomic source, beamsplitters and mirrors for a wide range of Bragg diffraction experiments and interferometers. For example, one could, in tangent with a few practical constraints, use our analysis to calculate the optimal velocity-selection pulse to use for a thermal source MZ interferometer. Given the potential to increase BEC flux, alongside other practical advantages of a narrow momentum width such as a lower susceptibility to wavefront distortions and the Coriolis effect, we believe Bose-condensed sources to be a viable candidate for the next generation of precision sensors based on Bragg pulse atom interferometry.   

\section*{Acknowledgments}
We would like to acknowledge P. A. Altin and G. McDonald for fruitful discussions, and the anonymous referees for their constructive criticisms. This work is supported by the Australian Research Council and the National Computational Infrastructure National Facility.

\section*{References}
\bibliographystyle{unsrt.bst}
\bibliography{Szigeti_bib}
\end{document}